\documentclass[twocolumn,twocolappendix]{aastex63} %linenumbers

\usepackage{chngcntr}
\usepackage{enumitem}
\usepackage{amsmath}
\usepackage{txfonts}
\usepackage{xcolor}
\usepackage{graphicx}
\usepackage{gensymb}
\usepackage{booktabs}
\usepackage{bm}
\usepackage{lipsum}
\usepackage{fnpct}
\usepackage{soul}
\setstcolor{red}
\usepackage{hyperref}

\AtBeginDocument{\mathcode`v=\varv}

\providecommand{\sorthelp}[1]{} % used with the Planck_bib.bib file

\newcommand{\HI}{\ifmmode \mathrm{\ion{H}{1}} \else \ion{H}{1} \fi}
\newcommand{\HII}{\ifmmode \mathrm{\ion{H}{2}} \else \ion{H}{2} \fi}
\newcommand{\CII}{\ifmmode \mathrm{\ion{C}{2}} \else \ion{C}{2} \fi}
\newcommand{\nh}{\ifmmode N_{{\mathrm{H}} \, \mathrm{I}} \else $N_{{\mathrm{H}} \, \mathrm{I}}$\fi}

\newcommand{\Planck}{\textit{Planck}}
\newcommand{\Radcliffe}{\textit{Radcliffe}}
\def\GHz{\ifmmode $\,GHz$\else \,GHz\fi}
\def\MJysr{\ifmmode \,$MJy\,sr\mo$\else \,MJy\,sr\mo\fi}
\def\microns{\ifmmode \,\mu$m$\else \,$\mu$m\fi}

\def\kms{\ifmmode $\,km\,s$^{-1}\else \,km\,s$^{-1}$\fi}

\defcitealias{marchal_2019}{M19}
\defcitealias{leike_2020}{L20}
\defcitealias{blagrave_dhigls:_2017}{B17}
\defcitealias{taank_2022}{T22}
\defcitealias{pelgrims_2020}{P20}
\defcitealias{meyer91}{M91}

\received{2022 May 4}
\revised{2022 Sept 29}
\accepted{2022 Nov 4}
\submitjournal{ApJ}

\shorttitle{On the origin of the North Celestial Pole Loop}
\shortauthors{Marchal \& Martin}

\graphicspath{{./}{figures/}}

\begin{document}

\title{On the origin of the North Celestial Pole Loop}

\correspondingauthor{Antoine Marchal}
\email{amarchal@cita.utoronto.ca}

\author[0000-0002-5501-232X]{Antoine Marchal}
\affiliation{Canadian Institute for Theoretical Astrophysics, University of Toronto, 60 St. George Street, Toronto, ON M5S 3H8, Canada}
\affiliation{Research School of Astronomy \& Astrophysics, Australian National University, Canberra ACT 2610 Australia}

\author[0000-0002-5236-3896]{Peter G. Martin}
\affiliation{Canadian Institute for Theoretical Astrophysics, University of Toronto, 60 St. George Street, Toronto, ON M5S 3H8, Canada}

\begin{abstract} 
    % context heading (optional)
    The North Celestial Pole Loop (NCPL) provides a unique laboratory for studying the early stage precursors of star formation.
    %, in particular the condensation of the neutral interstellar medium, and the formation of molecular gas.
    % aims heading (mandatory)
    Uncovering its origin is key to understanding the dynamical mechanisms that control the evolution of its contents.
    % methods heading (mandatory)
    In this study, we explore the 3D geometry and the dynamics of the NCPL using high-resolution dust extinction data and \HI\ data, respectively.
    % results heading (mandatory)
    We find that material toward Polaris and Ursa Major is distributed along a plane similarly oriented to the \Radcliffe\ wave. The Spider projected in between appears disconnected in 3D, a discontinuity in the loop shape. 
    We find that the elongated cavity that forms the inner part of the NCPL is a protrusion of the Local Bubble (LB) likely filled with warm (possibly hot) gas that passes through and goes beyond the location of the dense clouds. 
    %The cavity is oriented perpendicular to the \Radcliffe\ wave and the plane containing the NCPL.
    %
    An idealized model of the cavity as a prolate spheroid oriented toward the observer, reminiscent of the cylindrical model proposed by \citet{meyer91}, encompasses the protrusion and fits into arcs of warm \HI\ gas expanding laterally to it. %, possibly due to pressure gradients perpendicular to its long axis.
    As first argued by \citet{meyer91}, the non-spherical geometry of the cavity and the lack of OB stars interior to it disfavor an origin caused by a single point-like source of energy or multiple supernovae. Rather, the formation of the protrusion could be related to the propagation of warm gas from the LB into a pre-existing non-uniform medium in the lower halo, the topology of which was likely shaped by past star formation activity along the Local Arm.
\end{abstract}


\section{Introduction} \label{sec:intro}

The North Celestial Pole Loop \citep[NCPL,][hereafter M91]{heiles_1984,heiles_1989,meyer91}, first referred to as the Polar ridge by \cite{fejes_1973}, has been of interest not only because of its distinctive circular geometry on the sky and relatively high gas density for its position high above the Galactic plane, but also for the formation of molecules in individual clouds embedded within diffuse (dusty) \HI\ gas \citep[e.g., H$_2$CO, NH$_3$, CO, and \ion{C}{2}:][]{magnani_1985,heithausen_1987,mebold_1987,magnani_1988,grossmann_1990,skalidis_2021}. Among these clouds are some of the high latitude molecular clouds (HLCs) discovered by \cite{blitz_1984} and catalogued by \citet[][hereafter MBM]{magnani_1985}. These structures are mostly devoid of stars, making them useful sites in which to study the physical processes that control the state of the diffuse interstellar medium (ISM) before star formation occurs \citep{barr2010}.

Along the NCPL, two high-latitude cirrus complexes are observed. The Spider,
named for its prominent ``legs" emanating from a central ``body", is located at the ``top" of the loop (in Galactic coordinates, as used throughout this paper); it has been examined in complementary studies of thermal dust emission and molecules along with \HI \citep{barr2010,planck2011-7.12,martin_ghigls:_2015,blagrave_dhigls:_2017}. The Ursa Major complex, including MBM clouds 27-30, is located further east; extensive studies have provided valuable information on the kinematic, magnetic field, and turbulent structure of the diffuse ISM, and on the transition from the neutral atomic to the molecular phase \citep{myers_1995,pound_1997,mamd_2003,tritsis_2019,skalidis_2021}.

\subsection{Clues pointing to a dynamical process at the origin of the loop}
\label{subsec:clues}

\cite{myers_1995} reported that the line-of-sight (los) component of the magnetic field strength inferred from the Zeeman effect in the 21\,cm line in Ursa Major is remarkably strong (locally $|B_{\parallel}|=19\,\mu$G) and extends over 15\,pc\footnote{If at a distance of 150\,pc. But see the larger distance estimate in Section~\ref{sec:dust-extinction}.} with $|B_{\parallel}|>4\,\mu$G.\footnote{See also \citet{tritsis_2019,skalidis_2021} for complementary methods probing the plane-of-sky component.}  Enhancement of the magnetic field was also reported over larger scales along the entire loop by \cite{heiles_1989} using 13 Zeeman splitting measurements.  They found, in the cold neutral phase, $|B_{\parallel}|$ up to $12\,\mu$G and on average 5 times higher that the local Galactic magnetic field. In Ursa Major \cite{myers_1995} also found that the magnetic energy is comparable (within a factor 2) to the kinetic energy derived from \HI\ data, suggesting a coupling between the magnetic field and the gas motions. In the same region, \cite{pound_1997} reported a correlation along the filamentary structures (and not off the filaments) between the los component of the magnetic field strength and the column density inferred from IRAS 100\,$\mu$m images, suggesting that the correlation might originate from a dynamical process that compressed both the gas and field lines.

Additional early evidence of a dynamical process (or energetic event) associated with the NCPL was found by analyzing \HI\ data, radio continuum data at 408\,MHz, and soft X-rays data \citep{heiles_1974,haslam_1982,mccammon_1983}.
\citetalias{meyer91} found an enhancement of the count rates from soft X-rays in the central region of the NCPL, suggesting the presence of a cavity filled with hot plasma that could have initiated a lateral expansion. They noted a coincident minimum of emission at 408\,MHz, which if of synchrotron origin would require that the magnetic field had been expelled (consistent with the field enhancement observed along the loop). Based on the thick and thin model of Galactic non-thermal emission by \citet{beuermann_1985}, they calculated that the length of the radio continuum cavity would have to be 290\,pc to account for the radiation deficit.
This idea of an elongated expanding structure was bolstered by a strong velocity gradient from south east to north west seen in \HI\ data that could be explained by a model in which the NCPL is an expanding cylinder (of size about 30\,pc in radius and 180\,pc in length) oriented almost toward the observer (inclined 20\degree\ with respect to the Galactic plane), with central radial velocity -2\,\kms\ and lateral expansion about 20\,\kms.

Subsequently, motivated by the model proposed by \citetalias{meyer91}, a detailed study of the kinematics of molecular clouds in Ursa Major by \cite{pound_1997} using CO data showed that a model in which the clouds are located on the far side of an expanding bubble associated with the NCPL provides a good description of the data, including gas velocity, linewidth gradients, and the centroid velocity shift between atomic and molecular content. In this scenario, dense CO clumps are relics ``left behind" by a faster moving diffuse atomic gas carried along by the wind of the expanding bubble.

\subsection{An unclear origin of the dynamical process}

Although the aforementioned studies indicate a causal dynamical process for the loop, the origin of this process remains unclear. 

\citet{meyerdierks_1991b} found evidence that high velocity gas (i.e., HVC complex A, also called Chain A), intermediate, and low velocity gas in the NCPL region are related to each other, suggesting a cloud-Galaxy collision scenario. Despite such geometric and dynamic arguments, later studies have placed complex A at about $8-10$\,kpc \citep[][for the high-latitude part of complex A]{wakker_1996,ryans_1997b,van_Woerden_1999,wakker_2003,barger_2012}, ruling out this association.

\citet{heiles_1989} noted that the NCPL is unlikely to be a shell because very little gas is observed in its center, even at a different velocity, as would be characteristic of the expansion of a spherical structure caused by one or multiple supernovae explosions, as seen in the Galactic plane. \citet{heiles_1989} saw this structure rather as a large scale filament, in contrast to the subsequent model of an expanding cylinder proposed by \citetalias{meyer91}.
The non-spherical (cylindrical) model argues for a distributed source of energy, as opposed to a point-like source.
But \citetalias{meyer91} also noted that an origin from multiple supernovae and/or stellar winds is unlikely because no OB stars were observed within the loop \citep{humphreys_1978}. An association with OB stars also seems unlikely simply because of the high latitude.
More definitive empirical statements can now be attempted given a better distance to the loop structure (Section~\ref{subsec:dist}) and to OB stars using Gaia parallaxes and astrometry \citep{gonzalez_2021}; see Section~\ref{subsec:place}.

\subsection{Distance}
\label{subsec:dist}
The distance of the NCPL itself is a key to understanding its origin and the physical processes influencing its content. The first distance of a cirrus structure in the NCPL (the Ursa Major complex) was estimated to be roughly 100\,pc, uncertain by a factor of 2, by assuming that the cloud is part of the molecular material of the Galaxy with half-thickness about 70\,pc but located further than the extent of the local hot cavity around the Sun (now known as the LB) of size about 50\,pc \citep{deVries1987}.
This result was supported by \citet{penprase_1993} who found a distance to MBM 30 of $110\pm10$\,pc using interstellar absorption lines toward stars behind Ursa Major. 

However, the distance of these clouds have recently been re-visited using recent photometric surveys.
Using PanSTARRS photometry, \citet{schlafly_2014} found a much larger distance of about $350\pm25$\,pc for the Ursa Major MBM 30 cloud, placing it at about the same distance as Polaris. 
Using the 3D dust extinction map by \citet{green_2018}, \citet{tritsis_2019} found a distance of about 300\,pc toward the western part of the Ursa Major complex, consistent with recent polarimetric data presented in \citet{skalidis_2021}.
Using \textit{Gaia} DR2 data combined with PanSTARRS and 2MASS photometry, \cite{zucker_2019} found a distance of 369$^{+19}_{-22}\pm18$\,pc for the Spider and $371\pm3\pm19$\,pc for Ursa Major, respectively. The first and second $\pm$ values are statistical and systematic uncertainties, respectively.

In this work, we find that complexes in the NCPL, as well as the diffuse material that surrounds them, have distances varying from about 310\,pc to 450\,pc (Section~\ref{sec:dust-extinction}).

\subsection{Multiphase structure}
%Summarize Taank et al.
Using the Gaussian decomposition code {\tt ROHSA} \citep{marchal_2019}, \citet[][hereafter T22]{taank_2022} performed a spectral decomposition  of 21\,cm data from the GBT \HI\ Intermediate Galactic Latitude Survey (GHIGLS) and DRAO \HI\ Intermediate Galactic Latitude Survey (DHIGLS) surveys \citep[9\farcm4 and 1', respectively,][]{martin_ghigls:_2015,blagrave_dhigls:_2017}.
They identified four spatially (and dynamically) coherent components in the northern part of the NCPL, one of which is a remarkably well-defined warm arc \citepalias[called WNM$_{\rm A}$ in][]{taank_2022} moving at about 14\,\kms\ away from us, with a velocity dispersion of about 7.3\,\kms. The authors argued that this component is most likely a leftover from the dynamical event that caused the formation of cold gas within the loop.
Other components revealed that the cold and lukewarm phases together dominate the mass content of the neutral gas along the loop, in their region from Polaris east through the Spider to Ursa Major. Quantitative details of the respective phase mass fractions can be found in \citetalias{taank_2022}.
The spatial location of the moving warm arc and its relationship to the dense gas of the loop will be discussed in more detail in Section~\ref{sec:dynamics}.

\begin{figure}[!t]
    \centering
    \includegraphics[width=\linewidth]{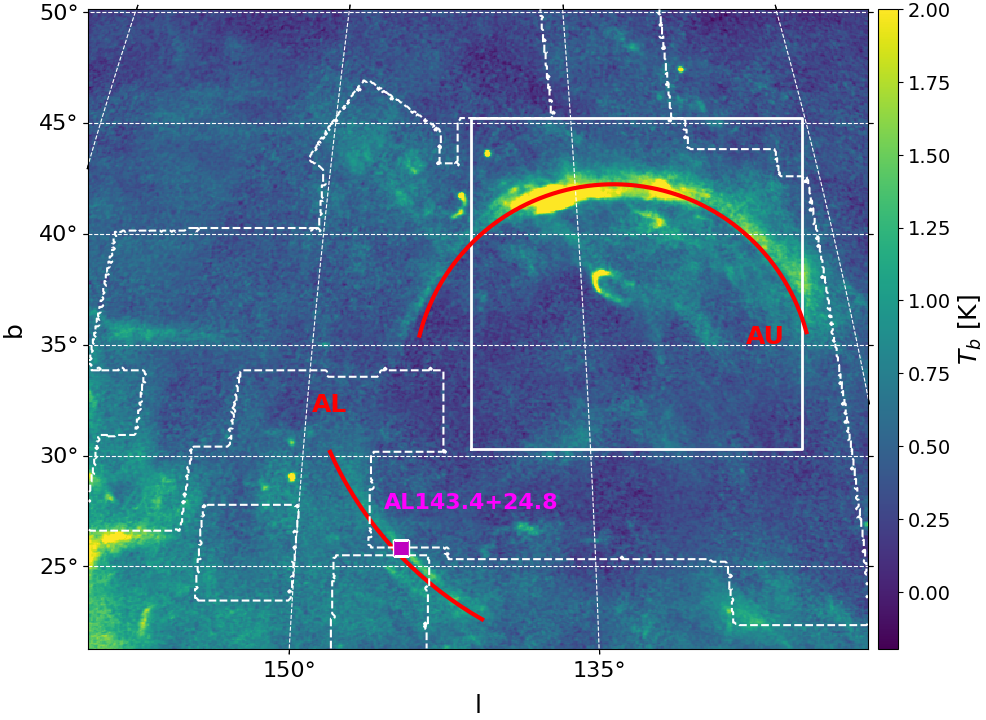}
    \includegraphics[width=\linewidth]{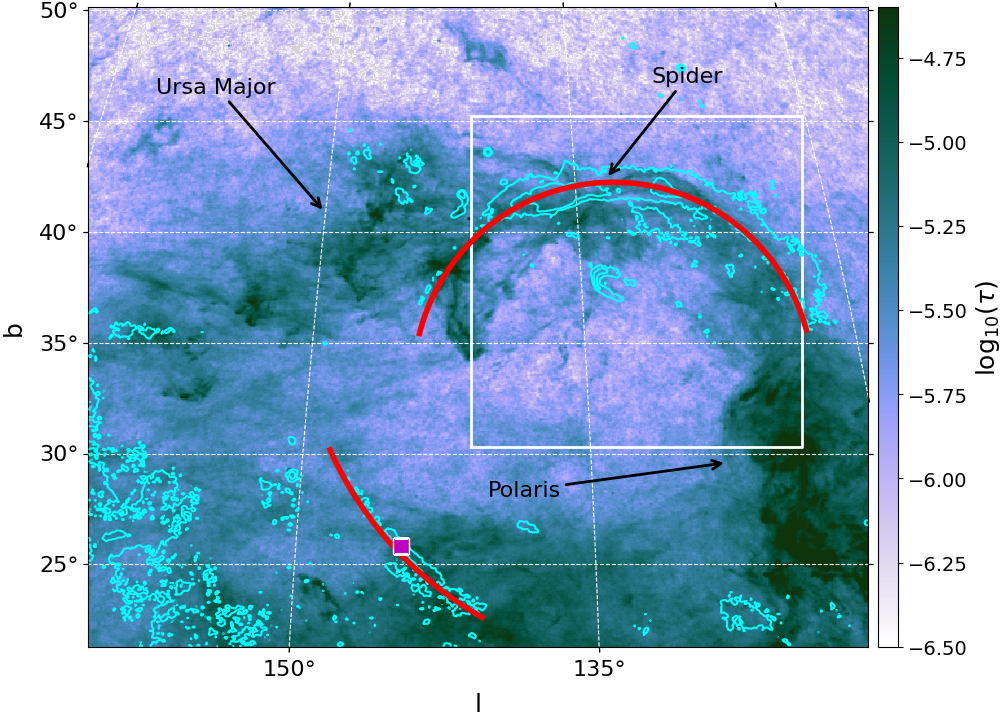}
    \caption{
    Top: Brightness temperature channel map of the NCPL region from EBHIS data at $v=16.85$\,\kms. 
    The white dashed outline shows the coverage of the NCPL mosaic from GHIGLS.
    The solid white box shows the 15\degree\ square region within which \citetalias{taank_2022} analysed the GHIGLS \HI\ data.
    Bottom: Dust optical depth map at 353\,GHz, $\tau_{353}$, from \cite{planck2013-p06b},
    with 1\,K and 2\,K contours of the brightness temperature map (top) overlaid in solid cyan.
    In both panels, red arcs show the location of the moving warm \HI\ gas and
    the magenta square shows the position of AL143.4+24.8.
    }
    \label{fig:GNILC-Model-Opacity_NCPL_mosaic}
\end{figure}

\subsection{Goal of this paper}
%Geometry of the NCPL with 3D tomography.
To add clarity to the uncertain situation previously outlined, toward understanding the origin of the NCPL in the large-scale context of the Solar neighborhood, we explore the 3D geometry of the loop using as high-resolution 3D dust extinction map and investigate its relationship to the moving \HI\ gas traced by the 21\,cm line.

%Organization paper
The paper is organized as follows.
% %
In Section~\ref{sec:dynamics} we present the \HI\ data used to probe the dynamics of the gas and discuss its relationship to the integrated dust optical depth traced by the \textit{Planck} satellite.
In Section~\ref{sec:dust-extinction}, we investigate the spatial distribution of interstellar medium (ISM) material using 3D dust tomography, assuming that dust and gas are well mixed.
A discussion of the place of the NCPL in the large scale context of the Solar neighborhood is presented in Section~\ref{sec:large-scale}.
Finally, a summary is provided in Section~\ref{sec:summary}.

A web page hosting the interactive figures of the paper is made available at \url{https://www.cita.utoronto.ca/NCPL/}.

\section{Moving \HI\ gas in the vicinity of the loop} 
\label{sec:dynamics}

\begin{figure}[!t]
    \centering
    \includegraphics[width=\linewidth]{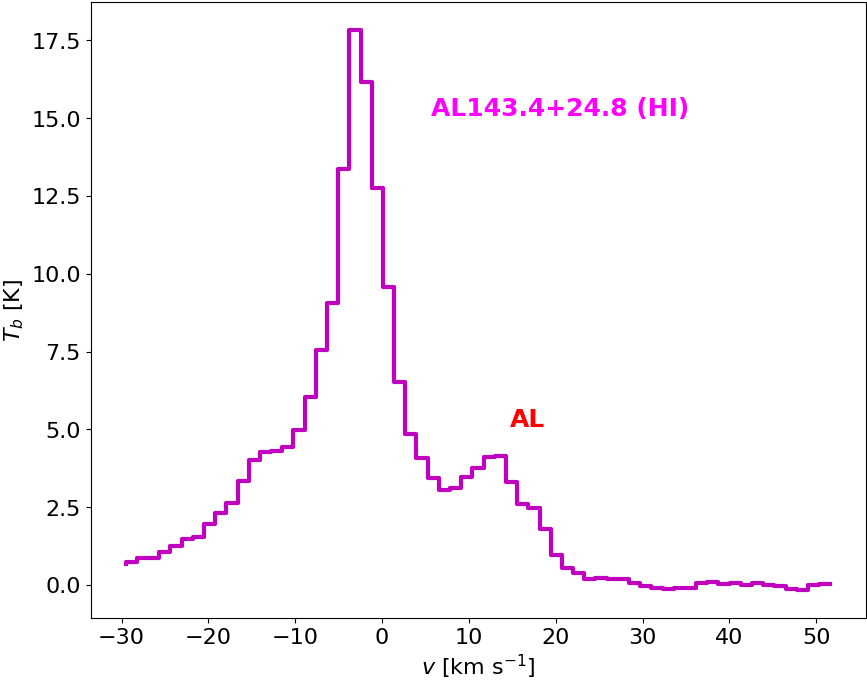}
    \caption{Brightness temperature profile along AL at ($143\fdg4$, $24\fdg8$) extracted from the NCPL mosaic in the GHIGLS survey.
    The emission at positive velocities corresponds to the AL arc seen in the brightness temperature map at $v=16.85$\,\kms\ shown in the top panel of Figure~\ref{fig:GNILC-Model-Opacity_NCPL_mosaic}}.
    \label{fig:Tb_spectra_NCPL_As}
\end{figure}

The top panel of Figure~\ref{fig:GNILC-Model-Opacity_NCPL_mosaic} shows the brightness temperature channel map of 21\, data from the Effelsberg-Bonn \HI\ Survey (EBHIS) \citep{kerp_2011,winkel_effelsberg-bonn_2016} at velocity $v=16.85$\,\kms\ that traces well the spatial location of the moving warm arc found by \citetalias{taank_2022}. The white dashed outline shows the coverage of the NCPL mosaic from the GHIGLS 21\,cm survey \citep{martin_ghigls:_2015} used by \citetalias{taank_2022} to probe the multiphase structure of the neutral gas within a $256 \times 256$ pixel (about 15\degree\ square) northern region denoted by the solid white box.

Outside the region analyzed by \citetalias{taank_2022}, a second smaller arc is seen at the same velocity, across the dust cavity (lower panel) in the south east part of the field.
In both panels, the locations of the larger ``upper" arc and the ``lower" arc are indicated by red circular arcs, which are hereafter called AU and AL, respectively.
Figure~\ref{fig:Tb_spectra_NCPL_As} shows an \HI\ brightness temperature line profile within AL, at ($143\fdg4$, $24\fdg8$),  annotated with a magenta square in Figure~\ref{fig:GNILC-Model-Opacity_NCPL_mosaic} and hereafter named AL143.4+24.8. From the broad feature at about 14\,\kms\ it can be appreciated that the gas is partly composed of the warm phase, similar to  AU or WNM$_{\rm A}$ \citepalias{taank_2022}, even without relying on a Gaussian decomposition.

The bottom panel of Figure~\ref{fig:GNILC-Model-Opacity_NCPL_mosaic} shows a map of the dust optical depth at 353\,GHz, $\tau_{353}$, from \cite{planck2013-p06b}, in the same projection and with the same annotations as in the top panel.
The locations of the Spider, Ursa Major, and Polaris are annotated with black arrows.
Solid cyan contours of the  1\,K and 2\, K \HI\ levels from the top panel probe the spatial relationship between AU, AL, and $\tau_{353}$.
AU follows the orientation of the northern part of the NCPL seen in dust, as discussed in \citetalias{taank_2022}. But for AL no similarly oriented structure is apparent in the dust map. Nevertheless, there is a significant optical depth at this location; whether this is due to foreground dust, dust associated with AL and the NCPL, or a combination of both will be discussed in Section~\ref{sec:dust-extinction}.

\section{Spatial distribution of dust along the line of sight}
\label{sec:dust-extinction}

At intermediate to high Galactic latitudes, the velocity of \HI\ cannot be used to estimate a kinematic distance. Instead we rely on 3D dust tomography, and the assumption that gas and dust are well mixed, to establish constraints on the distance and depth of the gaseous structures that make up the NCPL.

%  details of cube
We made use of the 3D mean\footnote{Here the mean refers to the average of the twelve samples generated, publicly available on Zenodo \citep[][{\tt mean\_std.h5}, doi:\url{10.5281/zenodo.3993082}]{reimar_leike_2020_3993082}} dust extinction density from \citet[hereafter L20]{leike_2020}.
\citetalias{leike_2020} used Gaia DR2 parallax and $G$-band photometric data combined with 2MASS, Pan-STARRS, and ALLWISE photometry to quantify interstellar extinction in a Cartesian $740\times740\times540$\,pc$^3$ volume centered on the Sun, with 1\,pc distance spacing and an effective resolution of 2\,pc.
Increasing x, y, and z are toward the Galactic centre, $\ell = 90$\degree, and positive $b$, respectively.

% units
The data unit is the differential extinction (specifically, the optical depth per parsec (e-folds pc$^{-1}$) in the Gaia $G$-band, which is referred to as the extinction density $s_x$.
Following \citet{zucker_2021}\footnote{More details about this conversion can be found in section 2.1.1 of \citet{zucker_2021}.} and \citet{bialy_2021}, the $G$-band extinction density can be converted to hydrogen volume density assuming $A_{G}/N_{\rm H}=4\times10^{-22}$\,cm$^2$\,mag from \citet{draine_2009}, leading to 
\begin{equation}
n_{\rm H}=880\, s_x\, {\mathrm{cm}}^{-3}\, , 
\label{eq:convert}
\end{equation}
where subscript ``H" denotes the total number of hydrogen nucleons, whether in the form of \HI, \HII, or H$_2$.

% profiles and PPV cube
We extracted individual extinction density profiles from the center (Sun's position) outward through the Cartesian cube in direction ($\ell, b$) using the {\tt dustmaps}\footnote{\url{https://dustmaps.readthedocs.io}. This tracks the ray through the cube and so is not quite equivalent to the extinction profile in a cone centered on the ray.} python package \citep{green_dustmaps_2018}. 
For any region of the sky, these profiles can be stored in a PPP cube, where the range of the third coordinate $d$ is limited by where the ray exits in the Cartesian cube. 

\begin{figure}[!t]
    \centering
    \includegraphics[width=0.91\linewidth]{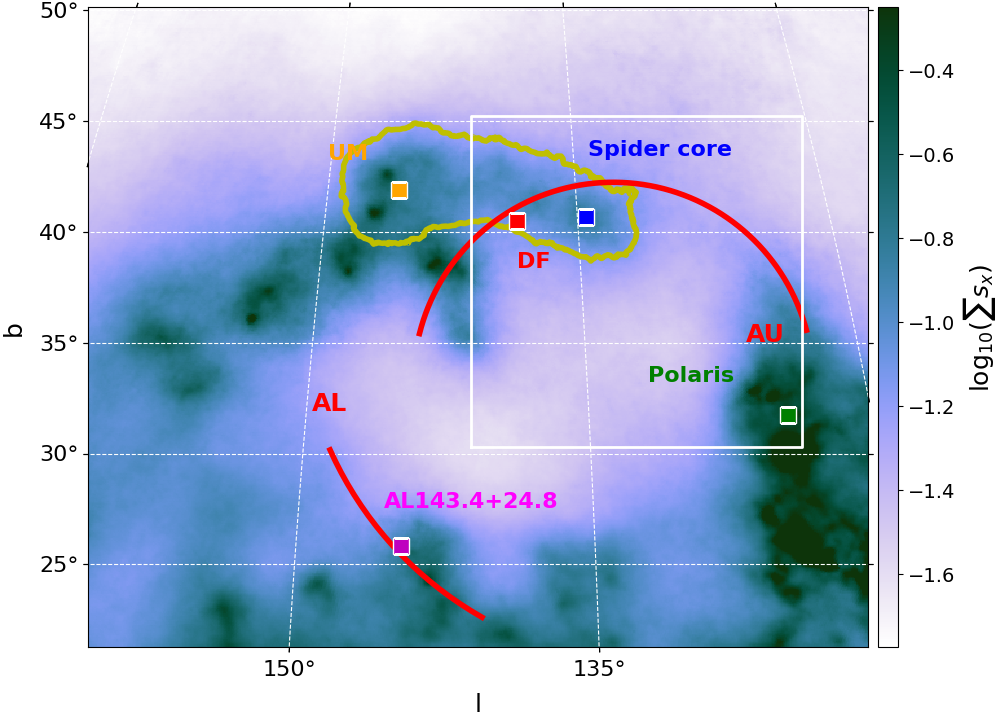}
    \includegraphics[width=0.91\linewidth]{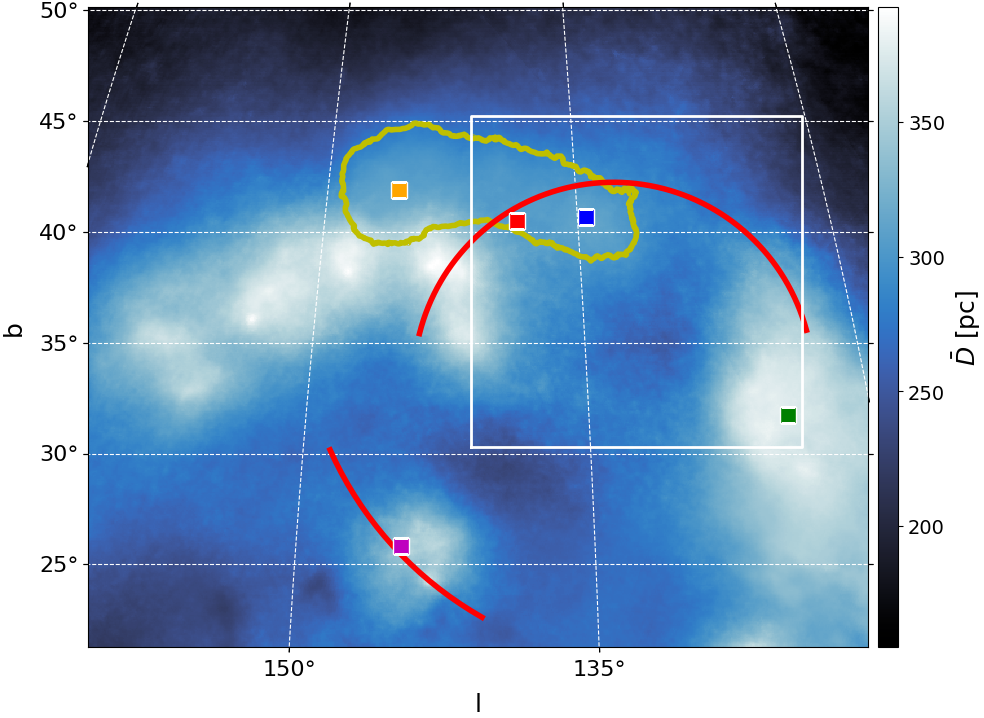}
    \includegraphics[width=0.91\linewidth]{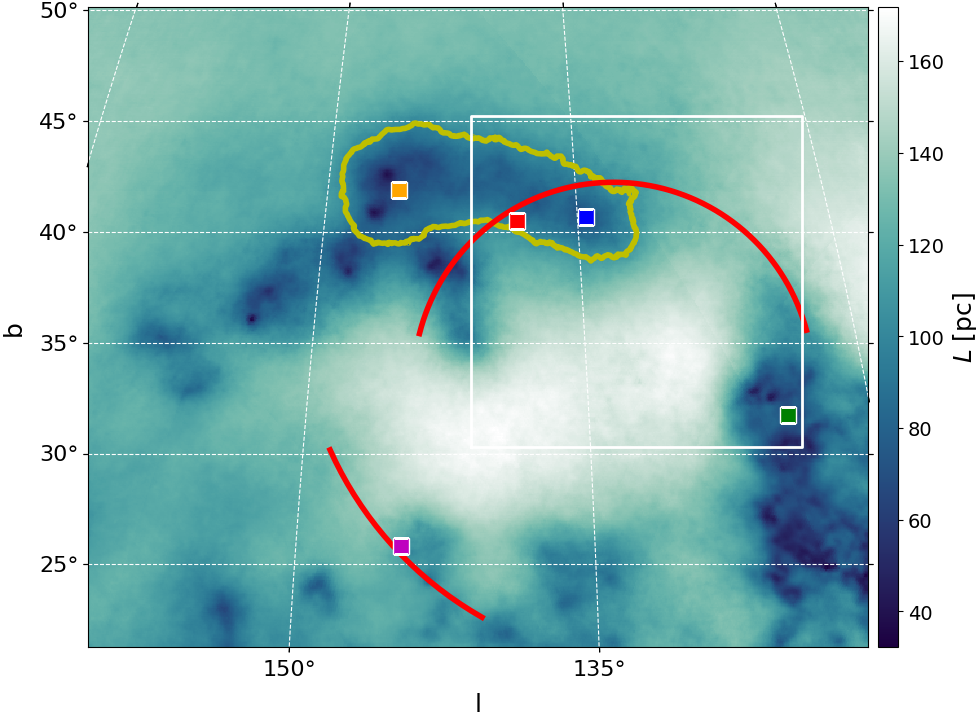}
    \caption{Integrated $G$-band extinction density (top) on a logarithmic scale), distance centroid (middle), and distance dispersion (bottom) from dust extinction profiles in the NCPL from the 3D dust extinction cube from \citetalias{leike_2020}. 
    The projection and annotations are as in Figure~\ref{fig:GNILC-Model-Opacity_NCPL_mosaic}. 
    The orange and red squares shows the positions of \HI\ absorption against background radio sources in the UM and DF regions in the DHIGLS survey. The blue, green, and magenta squares show lines of sight toward the Spider body and the northern part of Polaris, and AL143.4+24.8, respectively.
    The yellow contour at high Galactic latitude shows the coherent structure that includes the Spider and its extension to the east, and that in projection connects to Ursa Major and contributes to what is called the NCPL.
    }
    \label{fig:D_L_LEIKE_NCPL_mosaic}
\end{figure}

\subsection{Moment maps from the extinction density profiles}
Moment 0 of a profile is the integrated $s_x$ or simply the total extinction. For lines of sight (pixels) in the NCPL mosaic, this is shown on a logarithmic scale in the top panel of 
Figure~\ref{fig:D_L_LEIKE_NCPL_mosaic}, in the same projection and with the same annotations as in Figure~\ref{fig:GNILC-Model-Opacity_NCPL_mosaic}.
This shows the outline of the NCPL and the cavity below. Within the white box, the area studied in \citetalias{taank_2022}, the Spider body is a weak feature whose center is marked with a blue square.  Other specific directions discussed in Section~\ref{subsec:profiles} are marked according to Table~\ref{table:parameters-los}.

Moment 1, the distance centroid is shown in the middle
panel. The material in the extended NCPL to the east and west is the most distant and the Spider and AU appear to be closer. The material below the cavity and above NCPL in moment 0 is in the foreground, except for one region that coincides with the location of AL.

Moment 2, the distance dispersion, is shown in the bottom panel.
It can be interpreted as the typical depth of a dominant structure in extinction along the line of sight but if there are several components at different distances having similar extinctions, moment 2 overestimates the typical depth of the individual structures and is instead a measure of the distance between them. The NCPL stands out as having the lowest dispersion (greatest concentration of material along the line of sight) and by contrast the material in the cavity is more spread out along the line of sight. Note that high total extinction combined with low dispersion (high concentration along the line of sight) together imply localized high volume density, as seen directly in the Cartesian cube, and vice versa in the cavity.

The NCPL stands out in all three moment maps, though like in the \Planck\ map of $\tau_{353}$ does not follow AU faithfully. Toward the Spider the distance is about 320\,pc, closer than but consistent with estimate by \citet{zucker_2019} (Section~\ref{subsec:dist}).  In moments 0 and 2, there is a structure at the same distance as the Spider, extending to the northeast (traced fairly well by the blue, red, and orange fiducial markers, and shown by the yellow contour). This structure, with length, height, and depth of about 70, 15, and 80\,pc (see also Figure~\ref{fig:Av_LEIKE_second_quad} later on), is a clear localized concentration of denser material along the line of sight. 
There is cold and dense \HI\ gas seen in absorption against background radio sources at the positions of the red and orange squares. These two specific lines of sight (Table~\ref{table:parameters-los}) will hereafter be referred to as DF and UM, respectively, corresponding to the abbreviation of the DHIGLS fields within which they are embedded.

Material in the northern part of Polaris at the end of AU could be coherent with this, but there is a gap in between that has less extinction and is closer and less concentrated and less dense. Concentrations in moment 2 to the east in Ursa Major and in the west and south through Polaris represent more extinction and higher density and at 380\,pc are somewhat more distant than the Spider. The material toward AL143.4+24.8 is at a similar distance but is less concentrated.
The complex of different but adjoining dense structures (especially the Spider and its extension traced by the yellow contour) in projection contributes to what is called the NCPL.

\subsection{Extinction profiles along specific lines of sight}
\label{subsec:profiles}

\begin{deluxetable}{lccccccc}
\tablecaption{Basic properties of dust extinction profiles toward the NCPL}
\label{table:parameters-los}
\tablewidth{0pt}
\tablehead{
\colhead{Name} & \colhead{$l$} & \colhead{$b$} & \colhead{$N$}  & \colhead{$d_1$} & \colhead{$d_2$} & \colhead{$d_{\rm LB}$}\\
\nocolhead{} & \colhead{deg} & \colhead{deg} & & \colhead{pc} & \colhead{pc} & \colhead{pc} 
}
\startdata
$\color{red}\blacksquare\color{black}$ Deep Field (DF) & $138\fdg6$ & $40\fdg5$ & 2 & 320 & 400 & 286 \\
$\color{orange}\blacksquare\color{black}$ Ursa Major (UM) & $145\fdg7$ & $41\fdg9$ & 2 & 310 & 400 & 288 \\
$\color{blue}\blacksquare\color{black}$ Spider body & $134\fdg5$ & $40\fdg7$ & 1 & 320 & \dots &286 \\
$\color{green}\blacksquare\color{black}$ Polaris & $124\fdg5$ & $31\fdg7$ & 2 & 370& 450 &298 \\
$\color{magenta}\blacksquare\color{black}$ AL143.4+24.8 & $143\fdg4$ & $24\fdg8$ & 2 & 220 & 410 &230 \\
\enddata
\end{deluxetable}

\begin{figure}[!t]
    \centering
    \includegraphics[width=\linewidth]{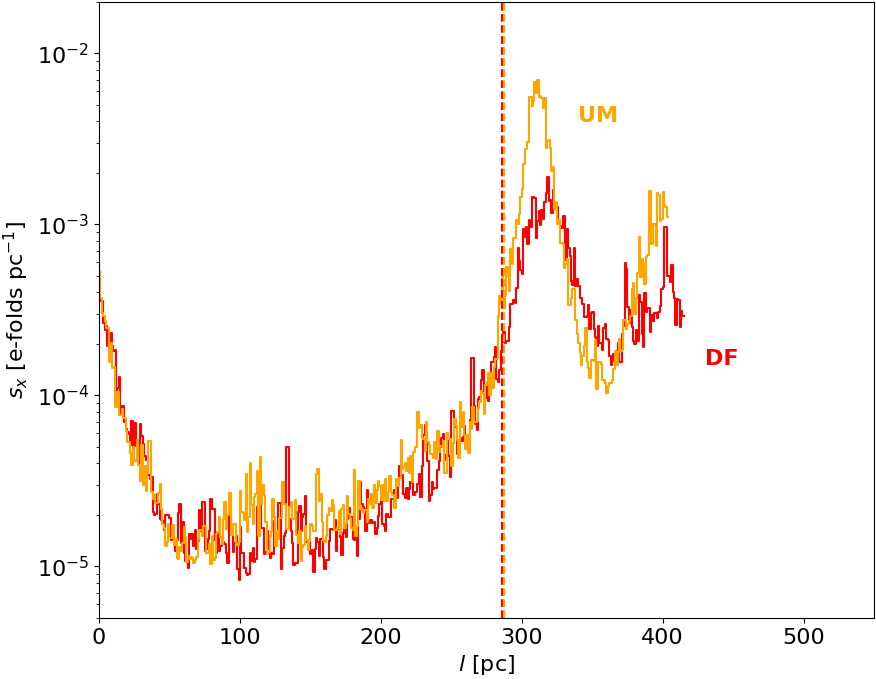}
    \includegraphics[width=\linewidth]{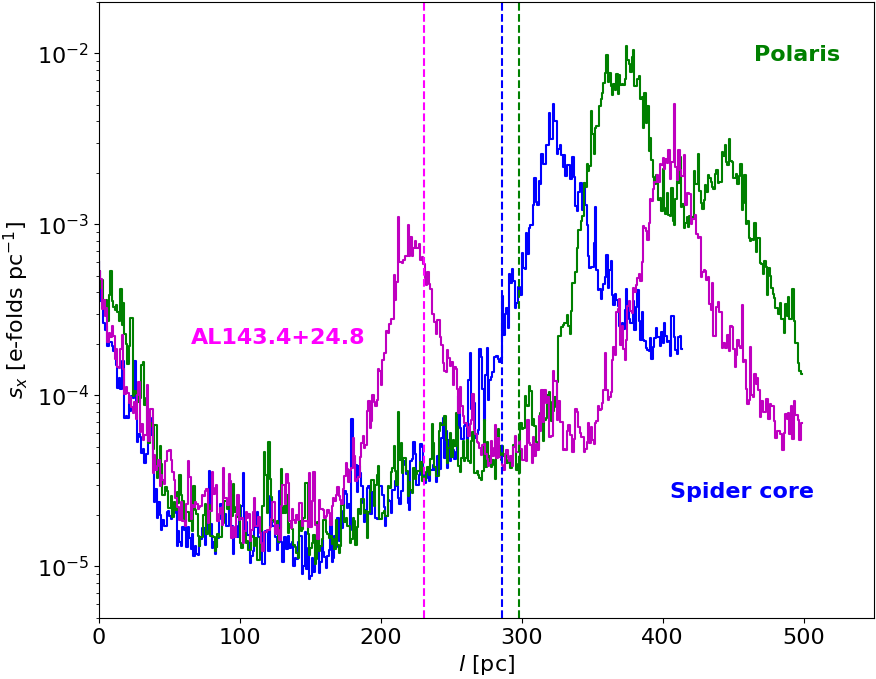}
    \caption{$G$-band extinction density in e-folds per pc along several lines of sight based on the 3D extinction cube from \citetalias{leike_2020}.  (Top): DF and UM lines of sight with \HI\ absorption measurements  and (Bottom): toward the Spider body, the northern part of Polaris, and AL143.4+24.8.
    Note that the sharp cuts at large distances are due to the finite coverage of the dust extinction model in the Cartesian cube, which varies with latitude.
    Vertical lines show the distance to the inner surface of the LB from
    \citetalias{pelgrims_2020}.
    }
    \label{fig:dust_spectra_NCPL}
\end{figure}

Figure~\ref{fig:dust_spectra_NCPL} shows individual dust extinction profiles along the DF (red) and UM (orange) lines of sight where there is \HI\ absorption from cool dense gas (top panel), and toward the Spider body (blue), the northern part of Polaris (green), and AL143.4+24.8 (magenta) in the smaller lower arc (bottom panel), respectively. Note the logarithmic scale. 
Basics properties of these five lines of sight are tabulated in Table~\ref{table:parameters-los}, including Galactic coordinates (also annotated in Figure~\ref{fig:D_L_LEIKE_NCPL_mosaic} using the same color code), number of structures along the line of sight (1 or 2), approximate distances to the identified structures $d_1$ and $d_2$, and the distance to the inner surface of the LB, $d_{\rm LB}$, as modeled by \citet[][their $l_{\rm max}=10$ model, hereafter P20]{pelgrims_2020}. 
Complementing Figure~\ref{fig:D_L_LEIKE_NCPL_mosaic}, the DF and UM, and the Spider profiles trace the same structure located at about 320\,pc. However, unlike toward the Spider body, the DF and UM profiles also show extinction from the extensive background complex in Ursa Major, at about 400\,pc. 
The latter provide interesting examples where two structures along the line of sight produce only a single peak in 21-cm emission and absorption \citepalias{taank_2022}. 
Note also that the closest structure in DF, UM, and the Spider body is adjacent to the inner surface of the LB, as expected from their operational definition of $d_{\rm LB}$, with a distance difference of about 30-40\,pc.

The profile toward the northern part of Polaris shows a double peak with the dominant structure at about 380\,pc and the second at about 450\,pc. In that direction, it is the dominant structure that determines the distance and dispersion in the moment 1 and 2 maps.
Finally, the profile toward AL143.4+24.8 shows two peaks as well. The first peak at about 220\,pc coincides with the inner surface of the LB (at 230\,pc)\footnote{However, here $d_{\rm LB}$ is not less than $d_1$, possible because the \citetalias{pelgrims_2020} model is smoothed by limiting the multipoles and/or because it uses different 3D extinction data from \citet{lallement_2019}.} and the second peak at about 410\,pc dominates in moment 1 with moment 2 being somewhat broadened by the closer peak.

\subsection{Distribution of low-density gas}

\begin{figure}[!t]
    \centering
    \includegraphics[width=\linewidth]{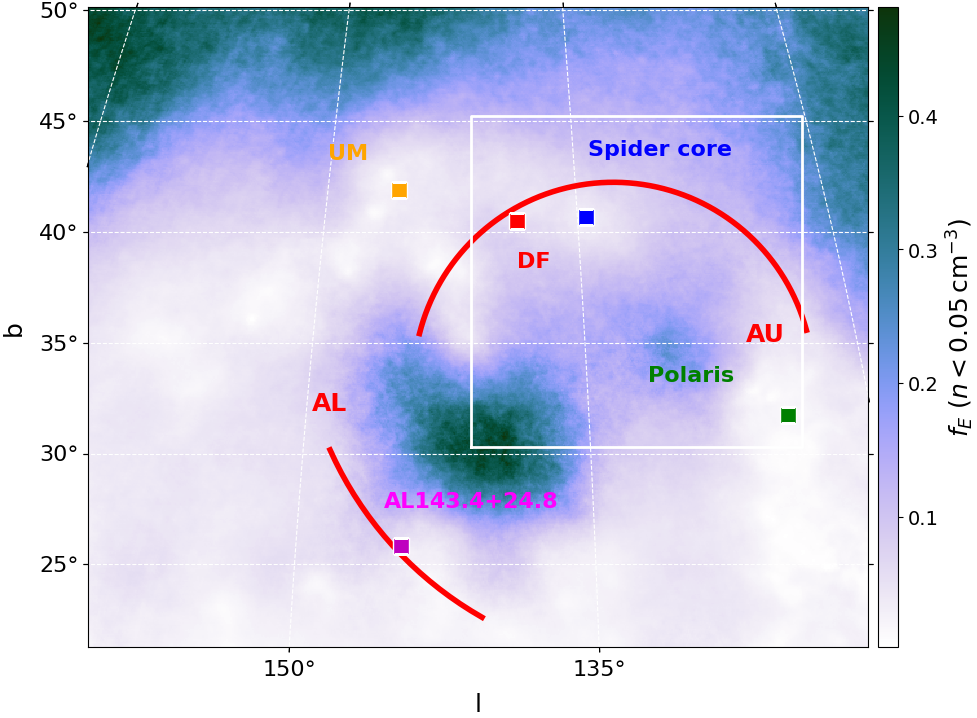}
    \includegraphics[width=\linewidth]{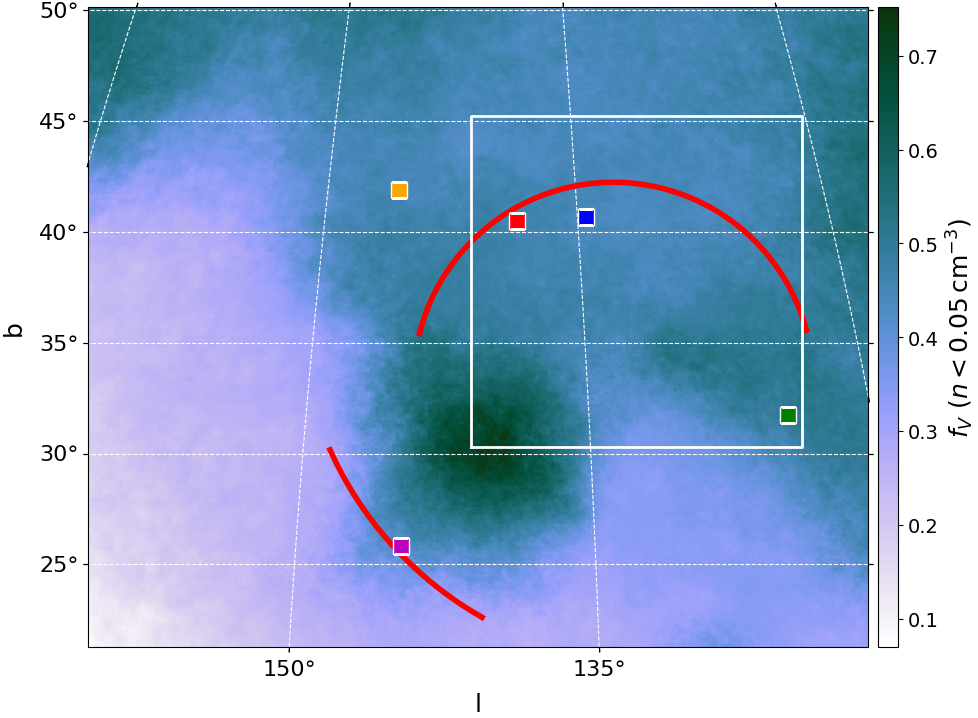}
    \caption{Fraction of total integrated $s_x$ of gas with $n_{\rm H}<0.05$\,cm$^{-3}$ (top). 
    Volume filling factor occupied by gas with $n_{\rm H}<0.05$\,cm$^{-3}$ (bottom). %
    The projection and the annotations are as in Figures~\ref{fig:GNILC-Model-Opacity_NCPL_mosaic} and \ref{fig:D_L_LEIKE_NCPL_mosaic}.
    Note that the color scale in the two panels corresponds to different ranges.
    }
    \label{fig:fv_NCPL}
\end{figure}

Focusing now on the low-density cavity, we explore the geometry of the material for which the dust $s_x$ indicates $n_{\rm H}<0.05$\,cm$^{-3}$ according to Equation~\ref{eq:convert}.
% this is the empirical part
For any line of sight from the Sun, we can find the fraction of the total volume that is occupied by such low density material ($f_V$), and the fractional contribution ($f_E$) to the total extinction shown above in Figure~\ref{fig:D_L_LEIKE_NCPL_mosaic} (top). These fractions are mapped for the NCPL region in 
Figure~\ref{fig:fv_NCPL} in the bottom and top panels, respectively.
The feature near the lower left corner of the white box has low total extinction (moment 0), high distance dispersion (moment 2), and high $f_V$ and $f_E$.\footnote{This feature is less prominent in the map of $\tau_{353}$, probably because \Planck\ probes dust along the entire line of sight out of the Galaxy, well beyond the range of the 3D extinction cube.}
Though less extreme, this combination is found in the low extinction the region enclosed by the arcs AU and AL. 

Low density material is spread out along the line of sight, which is reminiscent of the cylindrical geometry for the cavity found by \citetalias{meyer91}. As discussed using a 3D interactive plot in Section~\ref{subsec:vollow}, the cavity appears to be a protrusion of the LB.
Gas of such low density in the cavity could be in a warm neutral state (WNM, \citealp{wolfire_1995,wolfire_2003,bialy_2019}), an ionized state (WIM), a hotter plasma, or some combination. As discussed in Section~\ref{subsec:clues}, the hotter state is favored by the enhancement of the count rates from soft X-rays found by \citetalias{meyer91}.

By contrast, the previously discussed higher density concentrations along the loop have much lower $f_E$.

At higher latitude (above the loop), low-density gas also has high $f_E$ and $f_V$. This is expected from the known elongation of the LB cavity in the direction perpendicular to the Galactic plane \citep[at least to a distance higher than 250\,pc into the lower halo regions of both hemispheres,][]{lallement_2003}, also called the ``Local Chimney" \citep[hereafter LC,][]{welsh_1999}. As highlighted by \citet{lallement_2003}, for distances less than 400\,pc, no distinct and continuous neutral barrier to end the LC has been found in either Galactic hemisphere \citep{crawford_2002,welsh_2002}.

\subsection{Top view in Cartesian space}

\begin{figure*}
    \centering
    \includegraphics[width=0.49\linewidth]{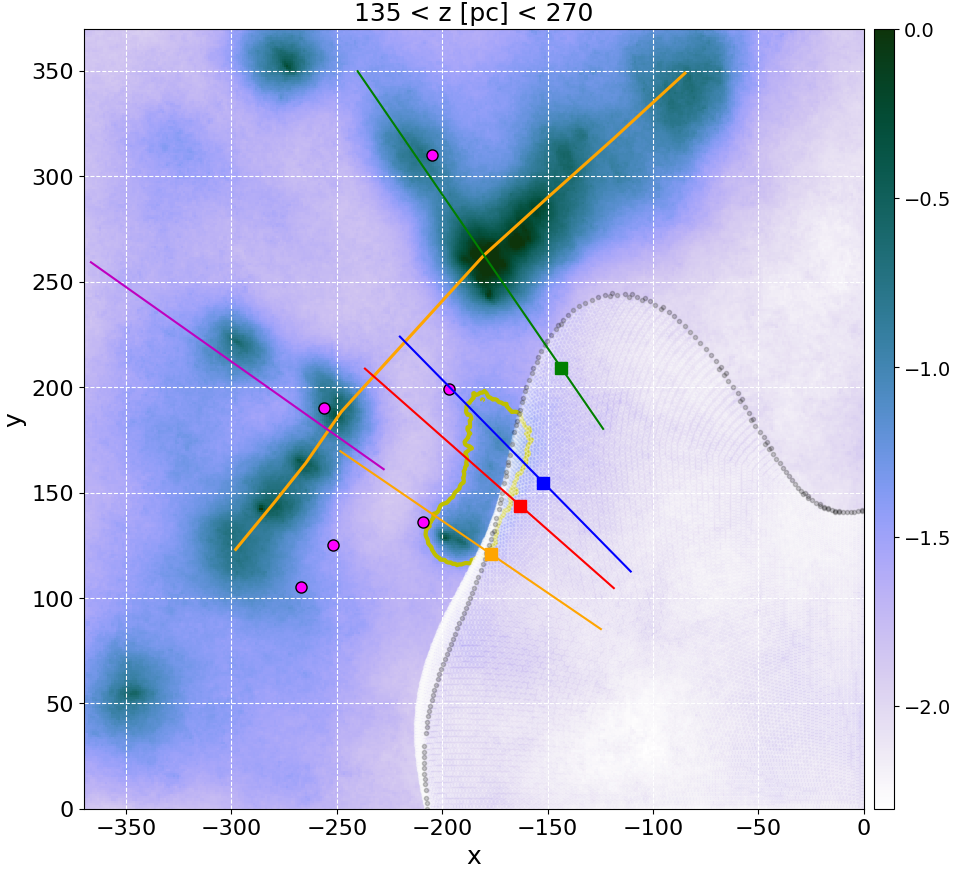}
    \includegraphics[width=0.49\linewidth]{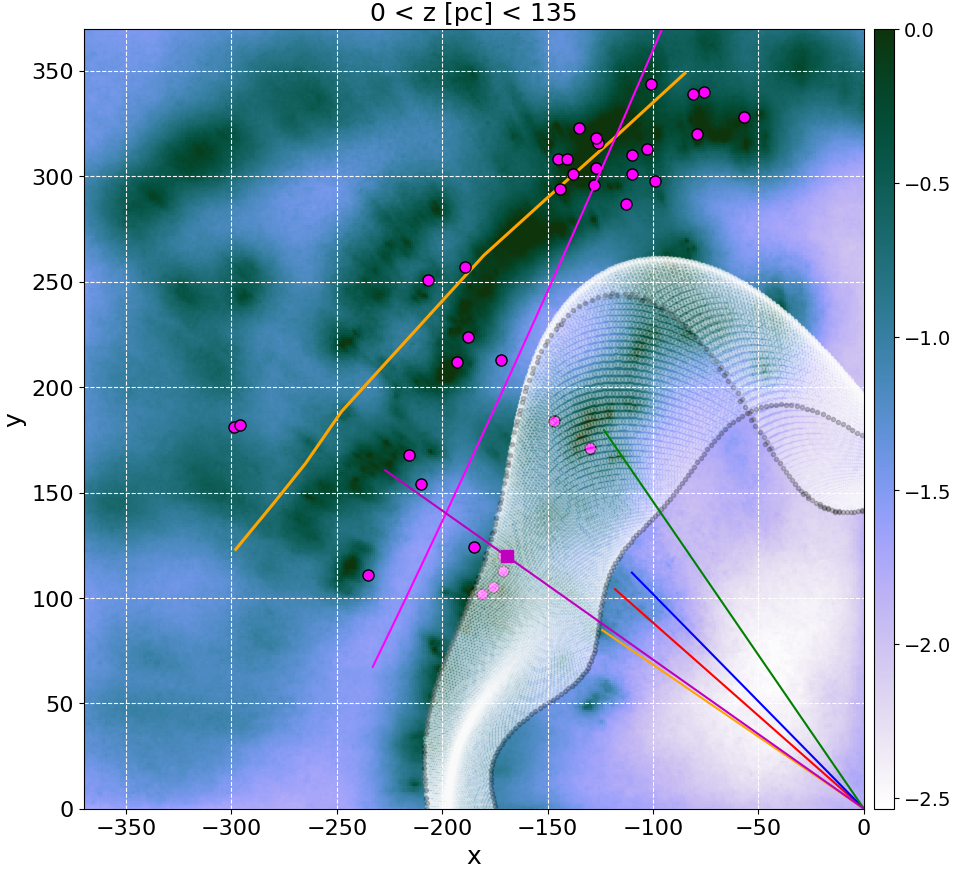}
    \caption{
    Top-down view of the second quadrant of $s_x$ integrated perpendicular to the Galactic plane (logarithmic form) for two layers above the plane: $135 < z < 270$\,pc (left) and $0 < z < 135$\,pc (right). The Sun is at (0,0) and the Galactic anti-center direction ($l=180\degree$) is toward the lower left corner.
    Fuchsia dots show the projected positions of molecular clouds in the layer of interest from \citet{zucker_2019}.
    The solid orange curve shows the projected orientation of the curved plane that includes material to the west of Polaris, the Polaris pillar, and Ursa Major.
    The yellow contour in the higher $z$ layer (left) shows the coherent structure that includes the Spider and its extension to the east located at $x=-190$\,pc, $y=130$\,pc, and that in projection on the sky connects to Ursa Major and contributes to what is called the NCPL.
    The solid magenta line in the bottom panel shows the projected position of the best-fit model of the \Radcliffe\ wave \citep{alves_2020}, which in this quadrant is above the Galactic plane in the lower layer.
    For orientation, projected path for the extinction density profiles from Figure~\ref{fig:dust_spectra_NCPL} are shown by solid lines on each panel in the volume of interest, using the same color code.
    The white drapery pattern shows projections of the inner surface of the \citetalias{pelgrims_2020} LB at various heights within each layer of interest; for orientation the projections are black for $z = 0$\,pc and $z = 135$\, pc.
    Colored squares show its intersection with the projected paths; these are the projections from the vertical-line positions shown in Figure~\ref{fig:dust_spectra_NCPL}.
    }
    \label{fig:Av_LEIKE_second_quad}
\end{figure*}

Moving to Cartesian space of the PPP cube, Figure~\ref{fig:Av_LEIKE_second_quad} shows a top-down view of the second quadrant of $s_x$ integrated perpendicular to the Galactic plane for two layers above the plane: $0 < z < 135$\,pc (left) and $135 < z < 270$\,pc (right). The Sun is at (0,0) and the Galactic anti-center direction ($l=180\degree$) is toward the lower left corner of the maps. 

The lower layer $0 < z < 135$\,pc \citep[about the HWHM of the cool gas layer,][and references within]{dickey_1990,kalberla_dark_2007,dickey_2021} contains large amounts of dust. Fuchsia dots show the projected positions of molecular clouds in the layer of interest from \citet{zucker_2019}. 
The magenta solid line shows the projected position of the best-fit model of the \Radcliffe\ wave, which in this quadrant is above the Galactic plane \citep{alves_2020}, and the white drapery pattern shows the \citetalias{pelgrims_2020} model of the inner surface of the LB in the layer of interest.

The upper layer $135 < z < 270$\,pc contains the structures of interest here. For orientation, projected paths of the $s_x$ profiles from Figure~\ref{fig:dust_spectra_NCPL} are shown by solid lines using the same color code.
In this view, the southern (lower layer) and northern (upper layer) parts of the Polaris ``pillar'' seen on the sky appear at the same distance from the Sun, indicating that the large extension of the pillar rises almost perpendicularly up to about 230 pc above the plane. Material to the west of Polaris, the Polaris pillar, and Ursa Major are distributed along a (slightly) curved plane, denoted by the orange curve.
At lower Galactic latitudes, the footprint of this plane seems close to the \Radcliffe\ wave, a 2.7-kpc-long filament of molecular gas that is hypothesized to be the molecular gas reservoir of the Local Arm (i.e., Orion-Cygnus arm) \citep{alves_2020,zucker_2022,swiggum_2022}. 

Interestingly, just as seen in Figure~\ref{fig:D_L_LEIKE_NCPL_mosaic}, the Spider and its extension to the east appear shifted relative to this plane, closer to us, at about the distance of the inner surface of the LB.
The yellow contour shows the coherent structure that includes the Spider and its extension to the east located at $x=-190$\,pc, $y=130$\,pc, and that in projection on the sky connects to Ursa Major.
There is also a tenuous link visible between Polaris and the Spider. In projection on the sky, the Spider, its extension, and this link contribute to the NCPL extending from Polaris to Ursa Major.

\begin{figure}[!t]
    \centering
    \includegraphics[width=\linewidth]{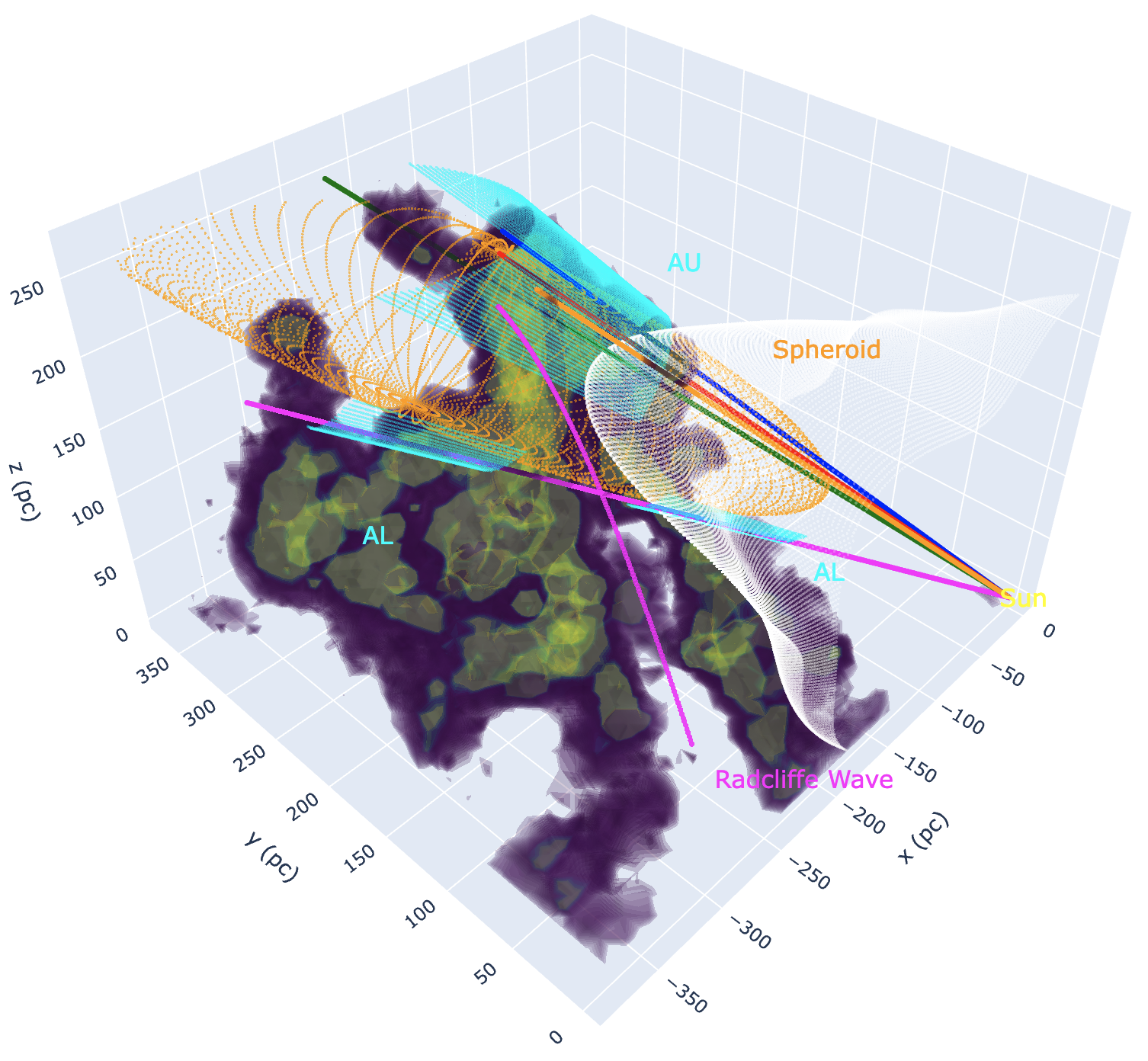}  
    \caption
    {
    \href{https://www.cita.utoronto.ca/NCPL/share/Marchal_Martin_2022/second_quad_above_plane.html}{Interactive} volume rendering of $s_x$ from \citetalias{leike_2020} in the second quadrant above the Galactic plane, highlighting intermediate-density gas 
    (logarithmic form, see text).
    As in Figure~\ref{fig:Av_LEIKE_second_quad}, for orientation, projected paths for the extinction density profiles from Figure~\ref{fig:dust_spectra_NCPL} are shown by color-coded solid lines.
    The white drapery pattern shows the inner surface of the LB \citepalias{pelgrims_2020}.
    The magenta curve shows the locus of the best-fit model of the \Radcliffe\ wave \citep{alves_2020}.    
    Cyan short cone segments show AL at $220\pm50$\,pc and $410\pm50$\,pc.
    The cyan long cone segment shows AU between the inner surface of the LB and the top layer of the cube.
    The orange prolate spheroid shows a simplified model of the cavity (see text).
    }
    \label{fig:3D_view_3b}
\end{figure}

\subsection{Interactive 3D view of the second quadrant above the Galactic plane}
\label{subsec:interactive_second}

\subsubsection{Volume rendering of intermediate-density gas}

Figure~\ref{fig:3D_view_3b} is a volume rendering of $s_x$ in the second quadrant above the Galactic plane, in logarithmic form between $-2.82$ and $-2.4$. According to Equation~\ref{eq:convert}, this highlights gas of intermediate density, in the range $1.3 \lesssim n_{\rm H}$ (cm$^{-3}) \lesssim 3.5$.
This has been made available online in \href{https://www.cita.utoronto.ca/NCPL/share/Marchal_Martin_2022/second_quad_above_plane.html}{interactive} form so that the interrelationships of various elements in 3D can be fully appreciated.\footnote{The rich interactive rendering, made with the plotly python library \citep{plotly}, might require tens of seconds to be loaded in a web browser and turning on and off individual overlaid structures several seconds.
The 3D renderings can be tumbled with the mouse, and it can be enlarged or shrunken by using the up- and down-scrolling buttons.}

As in Figure~\ref{fig:Av_LEIKE_second_quad}, paths of the $s_x$ profiles from Figure~\ref{fig:dust_spectra_NCPL} are shown by solid lines and the inner surface of the LB \citepalias{pelgrims_2020} is the white drapery pattern.

Cyan short cone segments show AL at $220\pm50$\,pc and $410\pm50$\,pc\footnote{The extent 50\,pc is arbitrarily chosen to cover a scale encompassing the dense and diffuse part of each structure.} (i.e., the two peaks seen in the $s_x$ profile; see also Table~\ref{table:parameters-los}), highlighting its two possible locations. 
The cyan long cone segment shows AU between the inner surface of the LB and the top layer of the extinction cube ($z=270$\,pc).

\begin{figure}[!t]
    \centering
    \includegraphics[width=\linewidth]{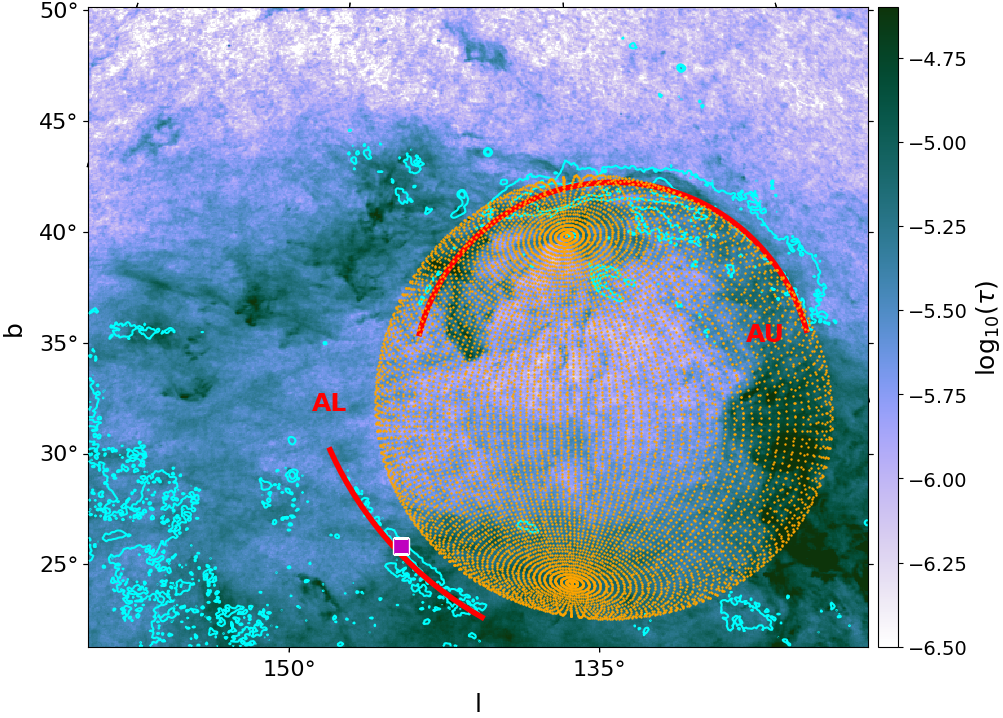}    
    \caption{
    Projected prolate spheroid (orange) overlaid on the dust optical depth map at 353\,GHz, $\tau_{353}$, from \cite{planck2013-p06b},
    with 1\,K and 2\,K contours of the brightness temperature map of the NCPL region from EBHIS data at $v=16.85$\,\kms\ overlaid in solid cyan.
    Red arcs show the location of the moving warm \HI\ gas.}\label{fig:HFI_CompMap_ThermalDustModel_2048_R1.20_NCPL_mosaic_model_spheroid}
\end{figure}

The orange prolate spheroid shown is an idealized model of the cavity, oriented approximately toward the observer.\footnote{Specifically, the spheroid is centred at centered at $x=-244$\,pc, $y=-236$\,pc, and $z=212$\,pc i.e., distance 400\,pc and ($\ell, b$) = (136\degree, 32\degree), and has semi-minor and semi-major axis of 55\,pc and 250\,pc, respectively.  Starting with its major axis oriented along the y axis, we applied a rotation of 31\degree\ around the x axis and 49\degree\ around the z axis. The inclination to the los through its center is about 3\degree.}
This model fits fairly well within AU and the AL. Note that it extends beyond the dust extinction cube. As seen in 3D, its surface touches the western part of Ursa Major, the Spider body and its extension northeastward, each structure seen along Polaris lines of sight, and the both structures along AL143.4+24.8.
Figure~\ref{fig:HFI_CompMap_ThermalDustModel_2048_R1.20_NCPL_mosaic_model_spheroid} shows a projected view of the spheroid overlaid on the dust optical depth map at 353\,GHz, $\tau_{353}$, from \cite{planck2013-p06b}, the image in Figure~\ref{fig:GNILC-Model-Opacity_NCPL_mosaic} (bottom), with similar contours and annotations.

The spatial proximity in 3D of AU and the farthest structure along AL143.4+24.8 (at about 410\,pc) favors a scenario in which the warm arc AL seen in \HI\ is also associated with the NCPL (located on the other side of the low dust extinction cavity). It remains possible that AL is spatially extended along the line of sight and traces the motion of warm neutral gas surrounding the elongated cavity (i.e., the prolate spheroid).

% %for on the sky context 

\begin{figure}[!t]
    \centering
    \includegraphics[width=\linewidth]{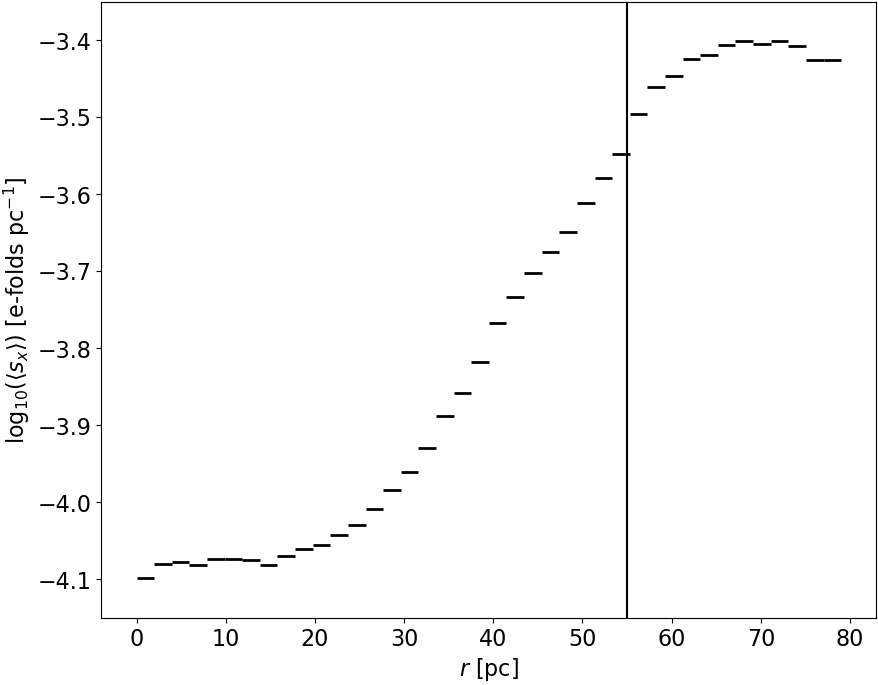}   
    \caption{Value of $\log_{10}$($\left<s_x\right>$) averaged within prolate spheroidal shells of fixed semi-major axis as a function of distance from the center measured along the semi-minor axis. The semi-minor axis of the orange spheroid in interactive Figure~\ref{fig:3D_view_3b} is annotated by the vertical line.}
    \label{fig:cavity_profile}
\end{figure}

This idealized model of a prolate spheroidal expanding cavity is reminiscent of both the expanding bubble proposed by \citet{pound_1997} and the cylindrical model proposed by \citetalias{meyer91}.
Figure~\ref{fig:cavity_profile} shows that the interior of the spheroid within the Cartesian 3D extinction cube is quite empty. For a fixed spheroid semi-major axis we have plotted \st{the average} $\log_{10}$($\left<s_x\right>$) averaged in spheroidal shells as the semi-minor axis is increased to slightly beyond that of the orange spheroid.

The \Radcliffe\ wave (magenta locus) and the plane that contains material to the west of Polaris, the Polaris pillar, and Ursa Major are similarly oriented, as can be appreciated in interactive Figure~\ref{fig:3D_view_3b}. Their relationship will be discussed in Section~\ref{sec:large-scale}.

\subsubsection{Volume rendering of low-density gas}
\label{subsec:vollow}

\begin{figure}[!t]
    \centering
    \includegraphics[width=\linewidth]{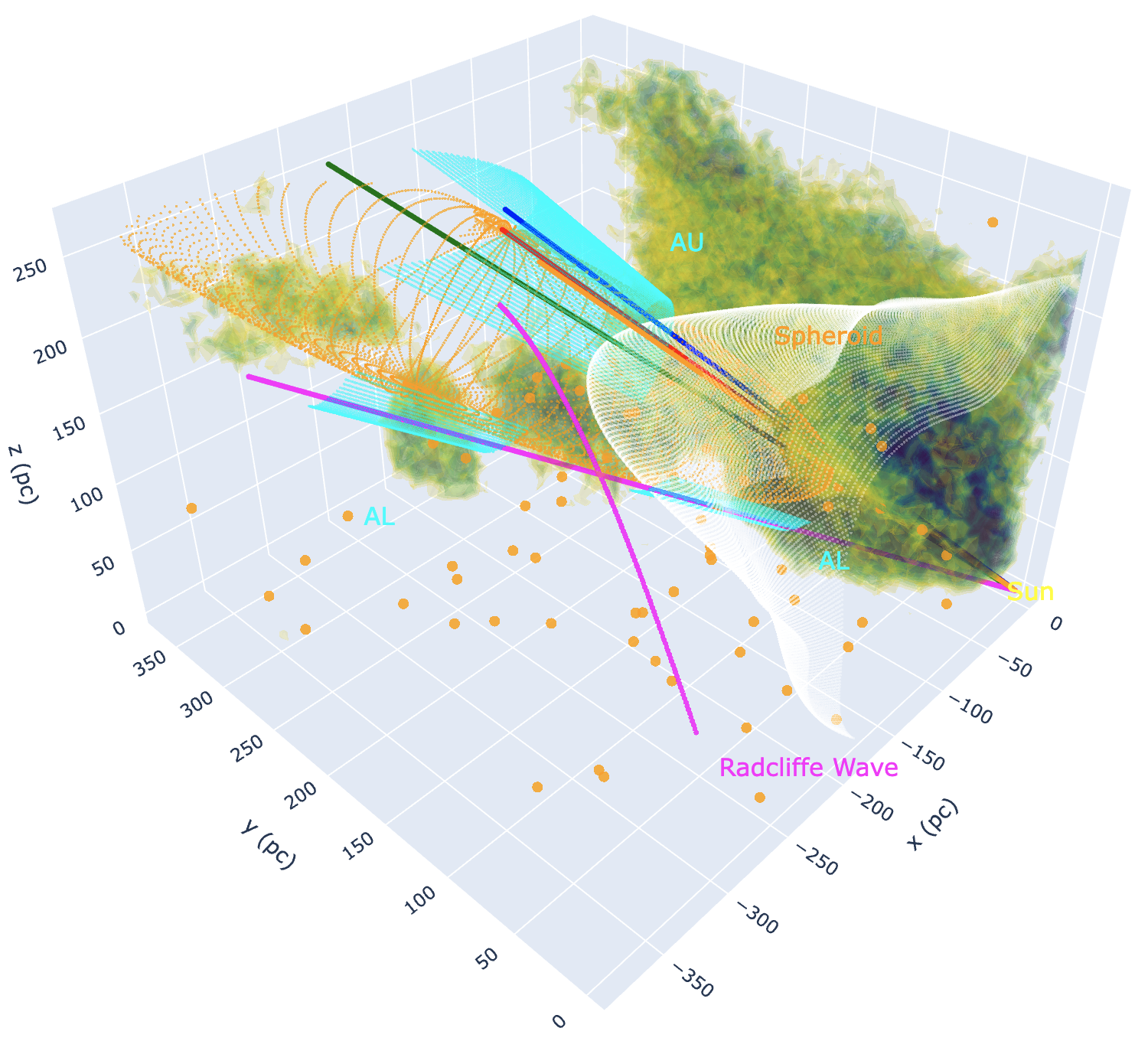}    
    \caption{\href{https://www.cita.utoronto.ca/NCPL/share/Marchal_Martin_2022/second_quad_above_plane_cut_dens_0.05.html}{Interactive} volume rendering of $s_x$ from \citetalias{leike_2020} in the second quadrant above the Galactic plane, as in Figure~\ref{fig:3D_view_3b} but highlighting low-density gas (see text).
    Annotations follow Figure~\ref{fig:3D_view_3b}.
    In addition, orange dots show the xyz positions of OB stars in this volume from \citet{gonzalez_2021} (Section~\ref{subsec:origin}).
    }
    \label{fig:3D_view_6}
\end{figure}

Figure~\ref{fig:3D_view_6} (\href{https://www.cita.utoronto.ca/NCPL/share/Marchal_Martin_2022/second_quad_above_plane_cut_dens_0.05.html}{interactive}) is a volume rendering of $s_x$ in the same quadrant, in logarithmic form between 
$-5.08$ and $-4.25$. This highlights low-density gas in the range $0.007 \lesssim n_{\rm H}$ (cm$^{-3}) \lesssim 0.05$.

The 3D distribution of low $s_x$ toward the cavity forms a protrusion of the LB that passes through and beyond the location of the cold higher-density clouds and reaches limit of the dust extinction cube (i.e., $z=270$\,pc). This is as inferred from the high volume filling factor of material with low $s_x$ (Figure~\ref{fig:fv_NCPL}).
Its geometry, well approximated by the interior of the orange prolate spheroid, is reminiscent of the cylindrical cavity found by \citetalias{meyer91}.
This protrusion filled with low $s_x$ is surrounded by the warm moving arcs that appear to have triggered the formation of cold gas in the NCPL \citepalias{taank_2022}.
Their dynamics support a scenario in which warm gas surrounding the protrusion expands laterally, presumably due to pressure gradients perpendicular to its long axis. 

%Added
Investigation of the mechanisms that control its lateral expansion, lead to the observed thermal condensation along the NCPL, and shape the gas (e.g., the prominent legs emanating from the central body of the Spider) will require numerical simulations. To choose the initial conditions for such an experiment, it is necessary to understand the place of the cavity in the large-scale context of the Solar neighborhood. This is motivated by the similar position and orientation of the \Radcliffe\ wave and the plane that contains material to the west of Polaris, the Polaris pillar, and Ursa Major.

\section{The NCPL in the larger-scale context of the Solar neighborhood}
\label{sec:large-scale}

% \begin{figure*}[!t]
%     \centering
%     \includegraphics[width=\linewidth]{figures/plotly_combined_label.png}   
%     % 
%     \caption{
%     Interactive volume renderings of $s_x$ from the entire \citetalias{leike_2020} 3D cube, highlighting gas in three limited ranges of density (logarithmic form, see text) in the \href{https://www.cita.utoronto.ca/NCPL/share/Marchal_Martin_2022/leike_full_higher_3.52.html}{top left},
%     \href{https://www.cita.utoronto.ca/NCPL/share/Marchal_Martin_2022/leike_full_1.32_3.52.html}{top right}, and
%     \href{https://www.cita.utoronto.ca/NCPL/share/Marchal_Martin_2022/leike_full_0.007_0.05.html}{bottom}.
%     %
%     For orientation, the path for the extinction density profile toward DF is shown by the red solid line.
%     %
%     The white drapery pattern the inner surface of the LB \citepalias{pelgrims_2020}.
%     %
%     The magenta locus and yellow line show the positions of the best-fit model of the \Radcliffe\ wave \citep{alves_2020} and the \textit{Split} \citep{lallement_2019,zucker_2022}, respectively.
%     %
%     The orange spheroid shows an idealized model of the cavity associated with the NCPL.
%     %
%     The large green sphere shows the Per-Tau shell \citep{bialy_2021}.
%     %
%     The two plum spheres show the \HI\ shells from \citet{bracco_2020}.
%     %
%     The small green sphere shows the surface of the Antlia SNR \citep{tetzlaff_2013,jung_2021}.
%     %
%     The small blue sphere shows the Vela SNR \citep{dodson_2003}.
%     }
%     \label{fig:3D_view_5}
% \end{figure*}

\begin{figure}[!t]
    \centering
    \includegraphics[width=\linewidth]{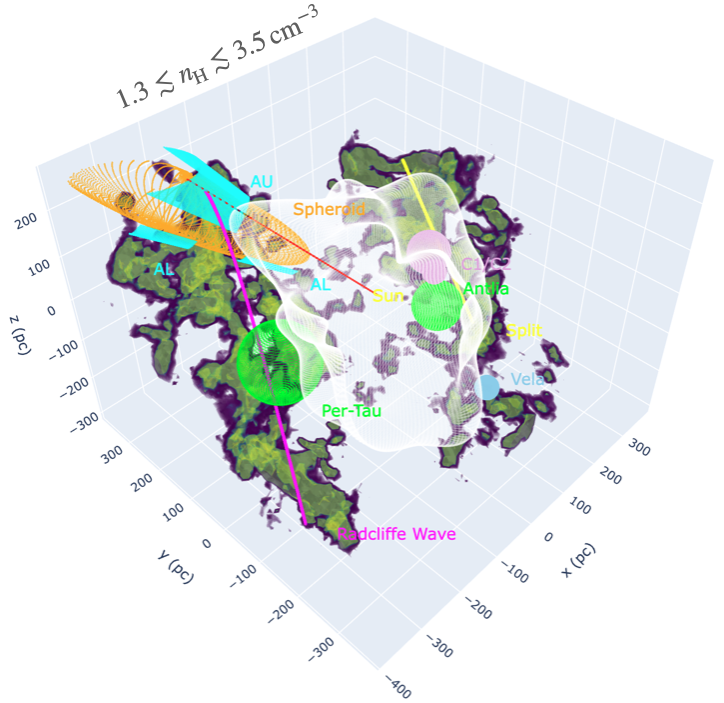}   
    \caption{
    \href{https://www.cita.utoronto.ca/NCPL/share/Marchal_Martin_2022/leike_full_1.32_3.52.html}{Interactive} volume renderings of $s_x$ from the entire \citetalias{leike_2020} 3D cube, highlighting gas in the limited range of density $1.3 \lesssim n_{\rm H}$ (cm$^{-3}) \lesssim 3.5$ (logarithmic form, see text).
    For orientation, the path for the extinction density profile toward DF is shown by the red solid line.
    The white drapery pattern the inner surface of the LB \citepalias{pelgrims_2020}.
    The magenta locus and yellow line show the positions of the best-fit model of the \Radcliffe\ wave \citep{alves_2020} and the \textit{Split} \citep{lallement_2019,zucker_2022}, respectively.
    The orange spheroid shows an idealized model of the cavity associated with the NCPL.
    The large green sphere shows the Per-Tau shell \citep{bialy_2021}.
    The two plum spheres show the \HI\ shells from \citet{bracco_2020}.
    The small green sphere shows the surface of the Antlia SNR \citep{tetzlaff_2013,jung_2021}.
    The small blue sphere shows the Vela SNR \citep{dodson_2003}.
    }
    \label{fig:mid_density}
\end{figure}

\begin{figure}[!t]
    \centering
    \includegraphics[width=\linewidth]{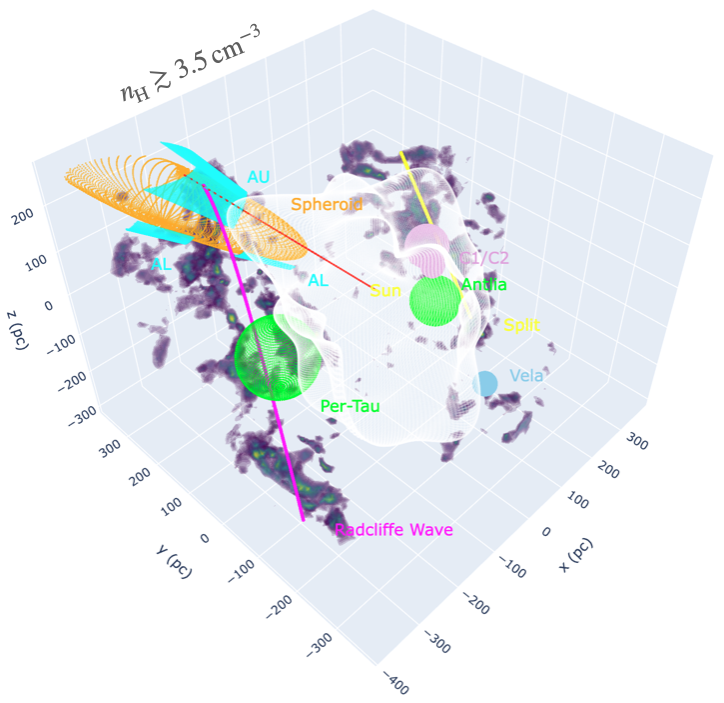}   
    \caption{
    \href{https://www.cita.utoronto.ca/NCPL/share/Marchal_Martin_2022/leike_full_higher_3.52.html}{Interactive} volume renderings of $s_x$ from the entire \citetalias{leike_2020} 3D cube, highlighting gas in the limited range of density $n_{\rm H} \gtrsim 3.5$ cm$^{-3}$ (logarithmic form, see text).
    Annotations are the same as in Figure~\ref{fig:mid_density}.
    }
    \label{fig:high_density}
    %fig:3D_view_5
\end{figure}

\begin{figure}[!t]
    \centering
    \includegraphics[width=\linewidth]{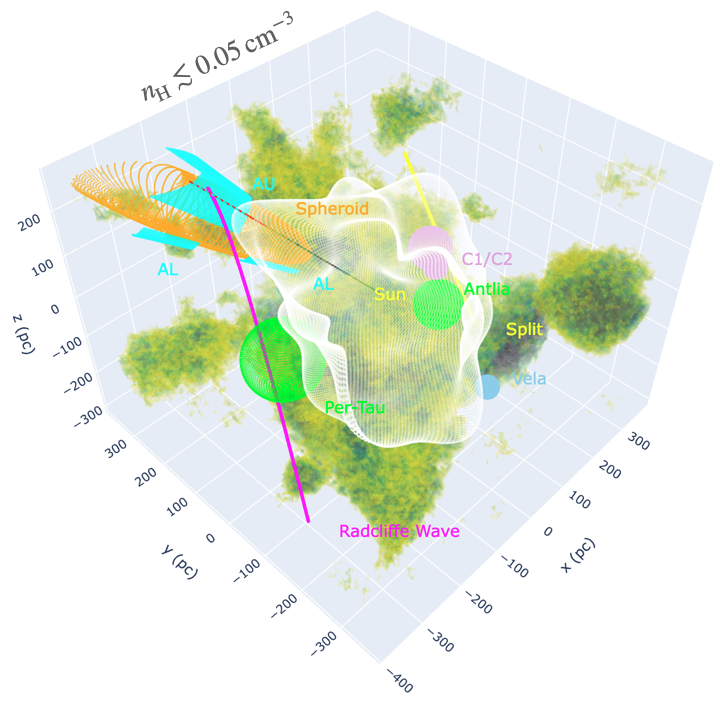}   
    \caption{
    \href{https://www.cita.utoronto.ca/NCPL/share/Marchal_Martin_2022/leike_full_0.007_0.05.html}{Interactive} volume renderings of $s_x$ from the entire \citetalias{leike_2020} 3D cube, highlighting gas in the limited range of density $n_{\rm H} \lesssim 0.05$ cm$^{-3}$(logarithmic form, see text).
    Annotations are the same as in Figure~\ref{fig:mid_density}.
    }
    \label{fig:low_density}
\end{figure}

\subsection{Place of the NCPL in the Solar neighborhood}
\label{subsec:place}

We investigated the place of the NCPL and its associated prolate spheroidal cavity,  the protrusion of the LB, in the larger-scale context of the Solar neighborhood, in all quadrants and above and below the Galactic plane.
Figure~\ref{fig:mid_density}
\href{https://www.cita.utoronto.ca/NCPL/share/Marchal_Martin_2022/leike_full_1.32_3.52.html}{interactive} shows a volume rendering of $s_x$ from \citetalias{leike_2020} in the whole $740\times740\times540$\,pc$^3$ volume around the Sun. In logarithmic form between -2.82 and -2.40, it highlights fairly dense gas in the limited range $1.3 \lesssim n_{\rm H}$ (cm$^{-3}) \lesssim 3.5$. 
Two other interactive renderings, for more dense and much less dense gas ($n_{\rm H} \gtrsim 3.5$ cm$^{-3}$ and $n_{\rm H} \lesssim 0.05$ cm$^{-3}$) are given in Figures~\ref{fig:high_density}
\href{https://www.cita.utoronto.ca/NCPL/share/Marchal_Martin_2022/leike_full_higher_3.52.html}{interactive} and \ref{fig:low_density}
\href{https://www.cita.utoronto.ca/NCPL/share/Marchal_Martin_2022/leike_full_0.007_0.05.html}{interactive}, respectively.

The familiar annotations from Figures~\ref{fig:3D_view_3b} and \ref{fig:3D_view_6} can be used to locate the NCPL.
In addition to the magenta locus of the \Radcliffe\ wave, the dense molecular inner part of the Local Arm, the yellow solid line shows where the 2\,kpc scale dusty structure called the \textit{Split} passes through this volume.  The \textit{Split} is argued to be a spur-like feature bridging the Local and Sagittarius-Carina arms \citep{lallement_2019,zucker_2022}.

%NCPL cavity
At this location along the Local Arm, the \Radcliffe\ wave is above the Galactic plane and reaches a height $z$ of about 100\,pc at the position of Cepheus \citep{alves_2020}. The center of the spheroidal cavity is 150\,pc away from the \Radcliffe\ wave at its shortest distance. Note that the overall distribution in the rendering of the material with fairly high $s_x$ (top right) is very similar to the distribution of the higher $s_x$ material (top left), and that there is a skeleton of denser material within the latter. This suggests that the molecular phase is enveloped by a gas of intermediate density, likely the cold and/or lukewarm (and unstable) neutral phase of the ISM. The geometry of this neutral gas with fairly high $s_x$ therefore resembles a wave pattern similar to the \Radcliffe\ wave. One could expect that the molecular gas that forms the \Radcliffe\ wave has originated from the condensation of cold neutral gas initiated by stellar feedback \citep{zucker_2022}, followed by the \HI-to-H$_2$ transition. In this picture, part of the condensed gas might still be in the neutral phase, at the outskirts of molecular clouds. The neutral counterpart of this wave seems more diffuse and extended, connecting low density material between clouds (e.g., along the NCPL), with a higher excursion from the Galactic plane.

\subsection{Qualitative comparison with other known cavities}
%Other cavities
Two other large-scale cavities are observed along the \Radcliffe\ wave in this volume of interest, but in contrast to the case of the protrusion of the LB identified above, \textit{direct} stellar feedback mechanisms have been identified.
%Peu-Tau
In green is shown the Per-Tau shell the surface of which the Perseus and Taurus molecular clouds are embedded. \citet{bialy_2021} argued that this shell was triggered by previous stellar and supernova feedback events between 6 and 22\,Myr ago. Its center is 90\,pc away from the \Radcliffe\ wave at its shortest distance.

% %Orion-Eradinus
Not shown here, the Orion-Eridanus superbubble \citep{heiles_1976,reynolds_1979,heiles_1999}, located between Orion and the LB, also has a prolate spheroidal shape 
(see \citealp{pon_2014,pon_2016} for a detailed Kompaneets model fitting of the cavity). It is thought to result from the combined effects of ionizing UV radiation, stellar winds, and a sequence of supernova explosions from the Orion OB association \citep[][and references within]{soler_2018}.

For completeness, we added known cavities in the direction opposite to the Local Arm, along the \textit{Split}. This includes two interacting \HI\ shells found by \citet{bracco_2020}, shown by the two plum spheres, and the Antlia supernovae remnant (SNR) \citep{tetzlaff_2013,jung_2021} and the Vela SNR \citep{dodson_2003}, shown by the small green and blue spheres, respectively.

\subsection{Origin of the NCPL}
\label{subsec:origin}
%Link to star formation theory of the porosity of the ISM.
%
Positions of OB stars from \citet{gonzalez_2021} are shown by the orange dots in interactive Figure~\ref{fig:3D_view_6}; there are none in the cavity. As first argued by \citetalias{meyer91}, the non-spherical geometry of the cavity and the lack of OB stars disfavor an origin caused by a single point-like source of energy or multiple supernovae.
The cavity appears to be a protrusion from the LB and so the origin could be linked to the propagation of warm (possibly hot) gas from the LB into a pre-existing non-uniform medium in the lower halo, the topology of which was likely shaped by past star formation activity along the Local Arm. 

In a gaseous disk affected by supernovae at a rate on the order of 1 per 50 years Galaxy wide, interconnected cavities and tunnels are expected from the growth of clusters and chains of connected supernova remnants \citep{cox_smith_1974}. \citet{cox_smith_1974} predicted that the resulting tunnel network would have low density, high temperature, and a very low magnetic field strength; the field properties seem consistent with the minimum of emission at 408\,MHz inside the cavity \citepalias{meyer91} and the enhancement of the magnetic field strength in complexes along the loop \citep{myers_1995,heiles_1989,pound_1997,tritsis_2019,skalidis_2021}.

The LB itself is irregular with tunnels branching off through the dense gas surrounding it, e.g., the ``Lupus tunnel" that connects the LB with the Loop I cavity, and the tunnel connecting the CMa void \citep{gry_1995} and the supershell GSH 238+00+09 \citep{heiles_1998} adjacent to Orion-Eridanus \citep{welsh_1991,welsh_1994,vergely_2001,lallement_2003}.

The NCPL cavity, a protrusion of the LB, seems to reach the top of the disc layer and could be a tunnel (or chimney) connecting the LB with the lower halo.
Its spatial coincidence with the Local Arm (and its hypothesized molecular reservoir, the \Radcliffe\ wave, at lower $z$)
supports a scenario in which past star formation activity along the arm has provided favorable conditions for its formation, making it a typical part of a tunnel network that is expected to fill half the interstellar medium by volume \citep{cox_smith_1974}.

\section{Summary} 
\label{sec:summary}

Our novel study of the 3D structure and origin of the North Celestial Pole Loop is based on a combination of 3D dust extinction and \HI\ data.
The main conclusions are as follows.

\begin{itemize}
    \itemsep-0.2em
    \item In \HI\ we identified a second warm arc (AL) that encompasses the low dust extinction cavity seen in the dust optical depth map from \cite{planck2013-p06b} and moves at a similar velocity (about 14\,km\,s$^{-1}$) to the warm arc (AU) found by \citetalias{taank_2022} that follows the northern part of the NCPL.
    \item The inferred 3D distribution of intermediate extinction density from \citetalias{leike_2020} reveals that Polaris and Ursa Major are distributed along a plane located at about 380\,pc and that the Spider (and its extension toward the east) is shifted with respect to this plane, 60\,pc closer to us, forming a discontinuity in the loop shape in 3D but not in projection.
    \item The 3D distribution of low extinction density toward the cavity reveals that the cavity is a protrusion of the LB that passes through and goes beyond the location of the cold dense clouds and reaches the limit of the dust extinction cube (i.e., $z=270$\,pc). Its geometry is reminiscent of the cylindrical cavity found by \citetalias{meyer91}.
    \item An idealized model of the cavity as a prolate spheroid oriented approximately toward the observer, with semi-minor and semi-major axis of 55\,pc and 250\,pc, respectively, centered at $x=-244$\,pc, $y=-236$\,pc, and $z=212$\,pc, fits within AU and AL. As seen in 3D, its surface touches the western part of Ursa Major, the Spider body and its extension northeastward, each structure seen along Polaris lines of sight, and the both structures along AL143.4+24.8 associated with AL.
    \item The warm moving arcs that seem to have triggered the formation of cold gas in the NCPL \citepalias{taank_2022} envelop the protrusion filled with low extinction density. Their dynamics suggest a scenario in which warm gas surrounding the protrusion expands laterally, possibly due to pressure gradients in the direction perpendicular to its long axis.
    \item The large scale distribution of gas in the the second quadrant above the Galactic plane reveals that the \Radcliffe\ wave and the plane that contains material to the west of Polaris, the Polaris pillar, and Ursa Major are similarly oriented. 
    On large scales, in the Solar neighborhood, the overall distribution in the rendering of the material with fairly high extinction density is very similar to the distribution of the higher extinction density material, and there is a skeleton of denser material within the latter. This suggests that the molecular phase is enveloped by a gas of intermediate density, likely the cold and/or lukewarm neutral phase of the ISM. The geometry of this neutral gas with fairly high extinction density therefore resembles a wave pattern similar to the \Radcliffe\ wave, but with more diffuse gas connecting high density material and with a higher excursion above the Galactic plane. The prolate spheroidal cavity is oriented perpendicular to it.
    \item As first argued by \citetalias{meyer91}, the non-spherical geometry of the cavity and the lack of OB stars internal to it disfavor an origin caused by a single point-like source of energy or multiple supernovae. Rather, the formation of the cavity (or protrusion) could be related to the propagation of warm  gas (possibly hot gas, as suggested by an enhancement of the count rates from soft X-rays inside the cavity found by \citetalias{meyer91}) from the LB into a pre-existing non-uniform medium in the lower halo, the topology of which was likely shaped by past star formation activity along the Local Arm.
\end{itemize}

\acknowledgments
We acknowledge support from the Natural Sciences and Engineering Research Council (NSERC) of Canada. 
This research has made use of the NASA Astrophysics Data System. 
Part of the visualization, exploration, and interpretation of data presented in this work was made possible using the glue visualization software, supported under NSF grant Nos. OAC1739657 and CDS\&E:AAG-1908419.
We thank A. Bracco, R. Klessen, C. Matzner, M.-A. Miville-Deschênes, and C. Zucker for enlightening conversations, and V. Pelgrims for providing us a Healpix map of the LB inner surface model used in this work.
We thank the referee, Alyssa Goodman, whose comments and suggestions have improved the clarity of presentation.

\software{AstroPy, a community-developed core Python package for Astronomy \citep{astropy_2013, astropy_2018}, CMasher \citep{velden_2020}, glue \citep{beaumont_2015}, NumPy \citep{van_der_walt_2011}, Matplotlib \citep{hunter_2007}, plotly \citep{plotly}, and} SciPy \citep{SciPy-NMeth}.

\bibliography{main}

\begin{thebibliography}{}
\expandafter\ifx\csname natexlab\endcsname\relax\def\natexlab#1{#1}\fi
\providecommand{\url}[1]{\href{#1}{#1}}
\providecommand{\dodoi}[1]{doi:~\href{http://doi.org/#1}{\nolinkurl{#1}}}
\providecommand{\doeprint}[1]{\href{http://ascl.net/#1}{\nolinkurl{http://ascl.net/#1}}}
\providecommand{\doarXiv}[1]{\href{https://arxiv.org/abs/#1}{\nolinkurl{https://arxiv.org/abs/#1}}}

\bibitem[{{Alves} {et~al.}(2020){Alves}, {Zucker}, {Goodman}, {Speagle},
  {Meingast}, {Robitaille}, {Finkbeiner}, {Schlafly}, \& {Green}}]{alves_2020}
{Alves}, J., {Zucker}, C., {Goodman}, A.~A., {et~al.} 2020, \nat, 578, 237,
  \dodoi{10.1038/s41586-019-1874-z}

\bibitem[{{Astropy Collaboration} {et~al.}(2013){Astropy Collaboration},
  {Robitaille}, {Tollerud}, {Greenfield}, {Droettboom}, {Bray}, {Aldcroft},
  {Davis}, {Ginsburg}, {Price-Whelan}, {Kerzendorf}, {Conley}, {Crighton},
  {Barbary}, {Muna}, {Ferguson}, {Grollier}, {Parikh}, {Nair}, {Unther},
  {Deil}, {Woillez}, {Conseil}, {Kramer}, {Turner}, {Singer}, {Fox}, {Weaver},
  {Zabalza}, {Edwards}, {Azalee Bostroem}, {Burke}, {Casey}, {Crawford},
  {Dencheva}, {Ely}, {Jenness}, {Labrie}, {Lim}, {Pierfederici}, {Pontzen},
  {Ptak}, {Refsdal}, {Servillat}, \& {Streicher}}]{astropy_2013}
{Astropy Collaboration}, {Robitaille}, T.~P., {Tollerud}, E.~J., {et~al.} 2013,
  \aap, 558, A33, \dodoi{10.1051/0004-6361/201322068}

\bibitem[{{Astropy Collaboration} {et~al.}(2018){Astropy Collaboration},
  {Price-Whelan}, {Sip{\H{o}}cz}, {G{\"u}nther}, {Lim}, {Crawford}, {Conseil},
  {Shupe}, {Craig}, {Dencheva}, {Ginsburg}, {Vand erPlas}, {Bradley},
  {P{\'e}rez-Su{\'a}rez}, {de Val-Borro}, {Aldcroft}, {Cruz}, {Robitaille},
  {Tollerud}, {Ardelean}, {Babej}, {Bach}, {Bachetti}, {Bakanov}, {Bamford},
  {Barentsen}, {Barmby}, {Baumbach}, {Berry}, {Biscani}, {Boquien}, {Bostroem},
  {Bouma}, {Brammer}, {Bray}, {Breytenbach}, {Buddelmeijer}, {Burke},
  {Calderone}, {Cano Rodr{\'\i}guez}, {Cara}, {Cardoso}, {Cheedella}, {Copin},
  {Corrales}, {Crichton}, {D'Avella}, {Deil}, {Depagne}, {Dietrich}, {Donath},
  {Droettboom}, {Earl}, {Erben}, {Fabbro}, {Ferreira}, {Finethy}, {Fox},
  {Garrison}, {Gibbons}, {Goldstein}, {Gommers}, {Greco}, {Greenfield},
  {Groener}, {Grollier}, {Hagen}, {Hirst}, {Homeier}, {Horton}, {Hosseinzadeh},
  {Hu}, {Hunkeler}, {Ivezi{\'c}}, {Jain}, {Jenness}, {Kanarek}, {Kendrew},
  {Kern}, {Kerzendorf}, {Khvalko}, {King}, {Kirkby}, {Kulkarni}, {Kumar},
  {Lee}, {Lenz}, {Littlefair}, {Ma}, {Macleod}, {Mastropietro}, {McCully},
  {Montagnac}, {Morris}, {Mueller}, {Mumford}, {Muna}, {Murphy}, {Nelson},
  {Nguyen}, {Ninan}, {N{\"o}the}, {Ogaz}, {Oh}, {Parejko}, {Parley}, {Pascual},
  {Patil}, {Patil}, {Plunkett}, {Prochaska}, {Rastogi}, {Reddy Janga},
  {Sabater}, {Sakurikar}, {Seifert}, {Sherbert}, {Sherwood-Taylor}, {Shih},
  {Sick}, {Silbiger}, {Singanamalla}, {Singer}, {Sladen}, {Sooley},
  {Sornarajah}, {Streicher}, {Teuben}, {Thomas}, {Tremblay}, {Turner},
  {Terr{\'o}n}, {van Kerkwijk}, {de la Vega}, {Watkins}, {Weaver}, {Whitmore},
  {Woillez}, {Zabalza}, \& {Astropy Contributors}}]{astropy_2018}
{Astropy Collaboration}, {Price-Whelan}, A.~M., {Sip{\H{o}}cz}, B.~M., {et~al.}
  2018, \aj, 156, 123, \dodoi{10.3847/1538-3881/aabc4f}

\bibitem[{{Barger} {et~al.}(2012){Barger}, {Haffner}, {Wakker}, {Hill},
  {Madsen}, \& {Duncan}}]{barger_2012}
{Barger}, K.~A., {Haffner}, L.~M., {Wakker}, B.~P., {et~al.} 2012, \apj, 761,
  145, \dodoi{10.1088/0004-637X/761/2/145}

\bibitem[{{Barriault} {et~al.}(2010){Barriault}, {Joncas}, {Falgarone},
  {Marshall}, {Heyer}, {Boulanger}, {Foster}, {Brunt}, {Miville-Desch{\^e}nes},
  {Blagrave}, {Kothes}, {Landecker}, {Martin}, {Scott}, {Stil}, \&
  {Taylor}}]{barr2010}
{Barriault}, L., {Joncas}, G., {Falgarone}, E., {et~al.} 2010, \mnras, 406,
  2713, \dodoi{10.1111/j.1365-2966.2010.16871.x}

\bibitem[{{Beaumont} {et~al.}(2015){Beaumont}, {Goodman}, \&
  {Greenfield}}]{beaumont_2015}
{Beaumont}, C., {Goodman}, A., \& {Greenfield}, P. 2015, in Astronomical
  Society of the Pacific Conference Series, Vol. 495, Astronomical Data
  Analysis Software an Systems XXIV (ADASS XXIV), ed. A.~R. {Taylor} \&
  E.~{Rosolowsky}, 101

\bibitem[{{Beuermann} {et~al.}(1985){Beuermann}, {Kanbach}, \&
  {Berkhuijsen}}]{beuermann_1985}
{Beuermann}, K., {Kanbach}, G., \& {Berkhuijsen}, E.~M. 1985, \aap, 153, 17

\bibitem[{{Bialy} \& {Sternberg}(2019)}]{bialy_2019}
{Bialy}, S., \& {Sternberg}, A. 2019, \apj, 881, 160,
  \dodoi{10.3847/1538-4357/ab2fd1}

\bibitem[{{Bialy} {et~al.}(2021){Bialy}, {Zucker}, {Goodman}, {Foley}, {Alves},
  {Semenov}, {Benjamin}, {Leike}, \& {En{\ss}lin}}]{bialy_2021}
{Bialy}, S., {Zucker}, C., {Goodman}, A., {et~al.} 2021, \apjl, 919, L5,
  \dodoi{10.3847/2041-8213/ac1f95}

\bibitem[{Blagrave {et~al.}(2017)Blagrave, Martin, Joncas, Kothes, Stil,
  Miville-Deschênes, Lockman, \& Taylor}]{blagrave_dhigls:_2017}
Blagrave, K., Martin, P.~G., Joncas, G., {et~al.} 2017, \apj, 834, 126,
  \dodoi{10.3847/1538-4357/834/2/126}

\bibitem[{{Blitz} {et~al.}(1984){Blitz}, {Magnani}, \& {Mundy}}]{blitz_1984}
{Blitz}, L., {Magnani}, L., \& {Mundy}, L. 1984, \apjl, 282, L9,
  \dodoi{10.1086/184293}

\bibitem[{{Bracco} {et~al.}(2020){Bracco}, {Bresnahan}, {Palmeirim},
  {Arzoumanian}, {Andr{\'e}}, {Ward-Thompson}, \& {Marchal}}]{bracco_2020}
{Bracco}, A., {Bresnahan}, D., {Palmeirim}, P., {et~al.} 2020, \aap, 644, A5,
  \dodoi{10.1051/0004-6361/202039282}

\bibitem[{{Cox} \& {Smith}(1974)}]{cox_smith_1974}
{Cox}, D.~P., \& {Smith}, B.~W. 1974, \apjl, 189, L105, \dodoi{10.1086/181476}

\bibitem[{{Crawford} {et~al.}(2002){Crawford}, {Lallement}, {Price}, {Sfeir},
  {Wakker}, \& {Welsh}}]{crawford_2002}
{Crawford}, I.~A., {Lallement}, R., {Price}, R.~J., {et~al.} 2002, \mnras, 337,
  720, \dodoi{10.1046/j.1365-8711.2002.05959.x}

\bibitem[{{de Vries} {et~al.}(1987){de Vries}, {Heithausen}, \&
  {Thaddeus}}]{deVries1987}
{de Vries}, H.~W., {Heithausen}, A., \& {Thaddeus}, P. 1987, \apj, 319, 723,
  \dodoi{10.1086/165492}

\bibitem[{{Dickey} {et~al.}(1990){Dickey}, {Hanson}, \& {Helou}}]{dickey_1990}
{Dickey}, J.~M., {Hanson}, M.~M., \& {Helou}, G. 1990, \apj, 352, 522,
  \dodoi{10.1086/168555}

\bibitem[{{Dickey} {et~al.}(2021){Dickey}, {Dempsey}, {Pingel},
  {McClure-Griffiths}, {Jameson}, {Dawson}, {D{\'e}nes}, {Clark}, {Leahy},
  {Lee}, {Miville-Desch{\^e}nes}, {Stanimirovi{\'c}}, {Tremblay}, \& {van
  Loon}}]{dickey_2021}
{Dickey}, J.~M., {Dempsey}, J.~M., {Pingel}, N.~M., {et~al.} 2021, arXiv
  e-prints, arXiv:2111.04545.
\newblock \doarXiv{2111.04545}

\bibitem[{{Dodson} {et~al.}(2003){Dodson}, {Legge}, {Reynolds}, \&
  {McCulloch}}]{dodson_2003}
{Dodson}, R., {Legge}, D., {Reynolds}, J.~E., \& {McCulloch}, P.~M. 2003, \apj,
  596, 1137, \dodoi{10.1086/378089}

\bibitem[{{Draine}(2009)}]{draine_2009}
{Draine}, B.~T. 2009, in Astronomical Society of the Pacific Conference Series,
  Vol. 414, Cosmic Dust - Near and Far, ed. T.~{Henning}, E.~{Gr{\"u}n}, \&
  J.~{Steinacker}, 453.
\newblock \doarXiv{0903.1658}

\bibitem[{{Fejes} \& {Wesselius}(1973)}]{fejes_1973}
{Fejes}, I., \& {Wesselius}, P.~R. 1973, \aap, 24, 1

\bibitem[{{Green}(2018)}]{green_dustmaps_2018}
{Green}, G. 2018, The Journal of Open Source Software, 3, 695,
  \dodoi{10.21105/joss.00695}

\bibitem[{{Green} {et~al.}(2018){Green}, {Schlafly}, {Finkbeiner}, {Rix},
  {Martin}, {Burgett}, {Draper}, {Flewelling}, {Hodapp}, {Kaiser}, {Kudritzki},
  {Magnier}, {Metcalfe}, {Tonry}, {Wainscoat}, \& {Waters}}]{green_2018}
{Green}, G.~M., {Schlafly}, E.~F., {Finkbeiner}, D., {et~al.} 2018, \mnras,
  478, 651, \dodoi{10.1093/mnras/sty1008}

\bibitem[{{Grossmann} {et~al.}(1990){Grossmann}, {Heithausen}, {Meyerdierks},
  \& {Mebold}}]{grossmann_1990}
{Grossmann}, V., {Heithausen}, A., {Meyerdierks}, H., \& {Mebold}, U. 1990,
  \aap, 240, 400

\bibitem[{{Gry} {et~al.}(1995){Gry}, {Lemonon}, {Vidal-Madjar}, {Lemoine}, \&
  {Ferlet}}]{gry_1995}
{Gry}, C., {Lemonon}, L., {Vidal-Madjar}, A., {Lemoine}, M., \& {Ferlet}, R.
  1995, \aap, 302, 497

\bibitem[{{Haslam} {et~al.}(1982){Haslam}, {Salter}, {Stoffel}, \&
  {Wilson}}]{haslam_1982}
{Haslam}, C.~G.~T., {Salter}, C.~J., {Stoffel}, H., \& {Wilson}, W.~E. 1982,
  \aaps, 47, 1

\bibitem[{{Heiles}(1976)}]{heiles_1976}
{Heiles}, C. 1976, \apjl, 208, L137, \dodoi{10.1086/182250}

\bibitem[{{Heiles}(1984)}]{heiles_1984}
---. 1984, \apjs, 55, 585, \dodoi{10.1086/190970}

\bibitem[{{Heiles}(1989)}]{heiles_1989}
---. 1989, \apj, 336, 808, \dodoi{10.1086/167051}

\bibitem[{{Heiles}(1998)}]{heiles_1998}
---. 1998, \apj, 498, 689, \dodoi{10.1086/305574}

\bibitem[{{Heiles} \& {Habing}(1974)}]{heiles_1974}
{Heiles}, C., \& {Habing}, H.~J. 1974, \aaps, 14, 1

\bibitem[{{Heiles} {et~al.}(1999){Heiles}, {Haffner}, \&
  {Reynolds}}]{heiles_1999}
{Heiles}, C., {Haffner}, L.~M., \& {Reynolds}, R.~J. 1999, in Astronomical
  Society of the Pacific Conference Series, Vol. 168, New Perspectives on the
  Interstellar Medium, ed. A.~R. {Taylor}, T.~L. {Landecker}, \& G.~{Joncas},
  211

\bibitem[{{Heithausen} {et~al.}(1987){Heithausen}, {Mebold}, \& {de
  Vries}}]{heithausen_1987}
{Heithausen}, A., {Mebold}, U., \& {de Vries}, H.~W. 1987, \aap, 179, 263

\bibitem[{{Humphreys}(1978)}]{humphreys_1978}
{Humphreys}, R.~M. 1978, \apjs, 38, 309, \dodoi{10.1086/190559}

\bibitem[{Hunter(2007)}]{hunter_2007}
Hunter, J.~D. 2007, CSE, 9, 90, \dodoi{10.1109/MCSE.2007.55}

\bibitem[{{Jung} {et~al.}(2021){Jung}, {McClure-Griffiths}, \&
  {Hill}}]{jung_2021}
{Jung}, S.~L., {McClure-Griffiths}, N.~M., \& {Hill}, A.~S. 2021, \mnras, 508,
  3921, \dodoi{10.1093/mnras/stab2773}

\bibitem[{Kalberla {et~al.}(2007)Kalberla, Dedes, Kerp, \&
  Haud}]{kalberla_dark_2007}
Kalberla, P. M.~W., Dedes, L., Kerp, J., \& Haud, U. 2007, Astronomy \&
  Astrophysics, 469, 511, \dodoi{10.1051/0004-6361:20066362}

\bibitem[{{Kerp} {et~al.}(2011){Kerp}, {Winkel}, {Ben Bekhti}, {Fl{\"o}er}, \&
  {Kalberla}}]{kerp_2011}
{Kerp}, J., {Winkel}, B., {Ben Bekhti}, N., {Fl{\"o}er}, L., \& {Kalberla},
  P.~M.~W. 2011, Astronomische Nachrichten, 332, 637,
  \dodoi{10.1002/asna.201011548}

\bibitem[{{Lallement} {et~al.}(2019){Lallement}, {Babusiaux}, {Vergely},
  {Katz}, {Arenou}, {Valette}, {Hottier}, \& {Capitanio}}]{lallement_2019}
{Lallement}, R., {Babusiaux}, C., {Vergely}, J.~L., {et~al.} 2019, \aap, 625,
  A135, \dodoi{10.1051/0004-6361/201834695}

\bibitem[{{Lallement} {et~al.}(2003){Lallement}, {Welsh}, {Vergely}, {Crifo},
  \& {Sfeir}}]{lallement_2003}
{Lallement}, R., {Welsh}, B.~Y., {Vergely}, J.~L., {Crifo}, F., \& {Sfeir}, D.
  2003, \aap, 411, 447, \dodoi{10.1051/0004-6361:20031214}

\bibitem[{Leike {et~al.}(2020)Leike, Glatzle, \&
  Enßlin}]{reimar_leike_2020_3993082}
Leike, R., Glatzle, M., \& Enßlin, T. 2020, {Galactic extinction within 400pc
  in cartesian coordinates}, 1.1,  Zenodo, \dodoi{10.5281/zenodo.3993082}

\bibitem[{{Leike} {et~al.}(2020){Leike}, {Glatzle}, \&
  {En{\ss}lin}}]{leike_2020}
{Leike}, R.~H., {Glatzle}, M., \& {En{\ss}lin}, T.~A. 2020, \aap, 639, A138,
  \dodoi{10.1051/0004-6361/202038169}

\bibitem[{{Magnani} {et~al.}(1985){Magnani}, {Blitz}, \&
  {Mundy}}]{magnani_1985}
{Magnani}, L., {Blitz}, L., \& {Mundy}, L. 1985, \apj, 295, 402,
  \dodoi{10.1086/163385}

\bibitem[{{Magnani} {et~al.}(1988){Magnani}, {Blitz}, \&
  {Wouterloot}}]{magnani_1988}
{Magnani}, L., {Blitz}, L., \& {Wouterloot}, J. G.~A. 1988, \apj, 326, 909,
  \dodoi{10.1086/166149}

\bibitem[{{Marchal} {et~al.}(2019){Marchal}, {Miville-Desch{\^e}nes}, {Orieux},
  {Gac}, {Soussen}, {Lesot}, {d'Allonnes}, \& {Salom{\'e}}}]{marchal_2019}
{Marchal}, A., {Miville-Desch{\^e}nes}, M.-A., {Orieux}, F., {et~al.} 2019,
  \aap, 626, A101, \dodoi{10.1051/0004-6361/201935335}

\bibitem[{Martin {et~al.}(2015)Martin, Blagrave, Lockman, Pinheiro~Gonçalves,
  Boothroyd, Joncas, Miville-Deschênes, \& Stephan}]{martin_ghigls:_2015}
Martin, P.~G., Blagrave, K. P.~M., Lockman, F.~J., {et~al.} 2015, \apj, 809,
  153, \dodoi{10.1088/0004-637X/809/2/153}

\bibitem[{{McCammon} {et~al.}(1983){McCammon}, {Burrows}, {Sanders}, \&
  {Kraushaar}}]{mccammon_1983}
{McCammon}, D., {Burrows}, D.~N., {Sanders}, W.~T., \& {Kraushaar}, W.~L. 1983,
  \apj, 269, 107, \dodoi{10.1086/161024}

\bibitem[{{Mebold} {et~al.}(1987){Mebold}, {Heithausen}, \&
  {Reif}}]{mebold_1987}
{Mebold}, U., {Heithausen}, A., \& {Reif}, K. 1987, \aap, 180, 213

\bibitem[{{Meyerdierks}(1991)}]{meyerdierks_1991b}
{Meyerdierks}, H. 1991, \aap, 251, 269

\bibitem[{{Meyerdierks} {et~al.}(1991){Meyerdierks}, {Heithausen}, \&
  {Reif}}]{meyer91}
{Meyerdierks}, H., {Heithausen}, A., \& {Reif}, K. 1991, \aap, 245, 247

\bibitem[{{Miville-Desch{\^e}nes} {et~al.}(2003){Miville-Desch{\^e}nes},
  {Joncas}, {Falgarone}, \& {Boulanger}}]{mamd_2003}
{Miville-Desch{\^e}nes}, M.~A., {Joncas}, G., {Falgarone}, E., \& {Boulanger},
  F. 2003, \aap, 411, 109, \dodoi{10.1051/0004-6361:20031297}

\bibitem[{{Myers} {et~al.}(1995){Myers}, {Goodman}, {Gusten}, \&
  {Heiles}}]{myers_1995}
{Myers}, P.~C., {Goodman}, A.~A., {Gusten}, R., \& {Heiles}, C. 1995, \apj,
  442, 177, \dodoi{10.1086/175433}

\bibitem[{{Pantaleoni Gonz{\'a}lez} {et~al.}(2021){Pantaleoni Gonz{\'a}lez},
  {Ma{\'\i}z Apell{\'a}niz}, {Barb{\'a}}, \& {Reed}}]{gonzalez_2021}
{Pantaleoni Gonz{\'a}lez}, M., {Ma{\'\i}z Apell{\'a}niz}, J., {Barb{\'a}},
  R.~H., \& {Reed}, B.~C. 2021, \mnras, 504, 2968,
  \dodoi{10.1093/mnras/stab688}

\bibitem[{{Pelgrims} {et~al.}(2020){Pelgrims}, {Ferri{\`e}re}, {Boulanger},
  {Lallement}, \& {Montier}}]{pelgrims_2020}
{Pelgrims}, V., {Ferri{\`e}re}, K., {Boulanger}, F., {Lallement}, R., \&
  {Montier}, L. 2020, \aap, 636, A17, \dodoi{10.1051/0004-6361/201937157}

\bibitem[{{Penprase}(1993)}]{penprase_1993}
{Penprase}, B.~E. 1993, \apjs, 88, 433, \dodoi{10.1086/191829}

\bibitem[{{\sorthelp{Planck Collaboration 2011X}}{Planck Collaboration
  XXIV}(2011)}]{planck2011-7.12}
{\sorthelp{Planck Collaboration 2011X}}{Planck Collaboration XXIV}. 2011, \aap,
  536, A24, \dodoi{10.1051/0004-6361/201116485}

\bibitem[{{\sorthelp{Planck Collaboration 2014K}}{Planck Collaboration
  XI}(2014)}]{planck2013-p06b}
{\sorthelp{Planck Collaboration 2014K}}{Planck Collaboration XI}. 2014, \aap,
  571, A11, \dodoi{10.1051/0004-6361/201323195}

\bibitem[{{Plotly Technologies Inc.}(2015)}]{plotly}
{Plotly Technologies Inc.} 2015, Collaborative data science,  Montreal, QC:
  Plotly Technologies Inc.
\newblock \url{https://plot.ly}

\bibitem[{{Pon} {et~al.}(2014){Pon}, {Johnstone}, {Bally}, \&
  {Heiles}}]{pon_2014}
{Pon}, A., {Johnstone}, D., {Bally}, J., \& {Heiles}, C. 2014, \mnras, 444,
  3657, \dodoi{10.1093/mnras/stu1704}

\bibitem[{{Pon} {et~al.}(2016){Pon}, {Ochsendorf}, {Alves}, {Bally}, {Basu}, \&
  {Tielens}}]{pon_2016}
{Pon}, A., {Ochsendorf}, B.~B., {Alves}, J., {et~al.} 2016, \apj, 827, 42,
  \dodoi{10.3847/0004-637X/827/1/42}

\bibitem[{{Pound} \& {Goodman}(1997)}]{pound_1997}
{Pound}, M.~W., \& {Goodman}, A.~A. 1997, \apj, 482, 334,
  \dodoi{10.1086/304136}

\bibitem[{{Reynolds} \& {Ogden}(1979)}]{reynolds_1979}
{Reynolds}, R.~J., \& {Ogden}, P.~M. 1979, \apj, 229, 942,
  \dodoi{10.1086/157028}

\bibitem[{{Ryans} {et~al.}(1997){Ryans}, {Keenan}, {Sembach}, \&
  {Davies}}]{ryans_1997b}
{Ryans}, R.~S.~I., {Keenan}, F.~P., {Sembach}, K.~R., \& {Davies}, R.~D. 1997,
  \mnras, 289, 986, \dodoi{10.1093/mnras/289.4.986}

\bibitem[{{Schlafly} {et~al.}(2014){Schlafly}, {Green}, {Finkbeiner}, {Rix},
  {Bell}, {Burgett}, {Chambers}, {Draper}, {Hodapp}, {Kaiser}, {Magnier},
  {Martin}, {Metcalfe}, {Price}, \& {Tonry}}]{schlafly_2014}
{Schlafly}, E.~F., {Green}, G., {Finkbeiner}, D.~P., {et~al.} 2014, \apj, 786,
  29, \dodoi{10.1088/0004-637X/786/1/29}

\bibitem[{{Skalidis} {et~al.}(2021){Skalidis}, {Tassis}, {Panopoulou},
  {Pineda}, {Gong}, {Mandarakas}, {Blinov}, {Kiehlmann}, \&
  {Kypriotakis}}]{skalidis_2021}
{Skalidis}, R., {Tassis}, K., {Panopoulou}, G.~V., {et~al.} 2021, arXiv
  e-prints, arXiv:2110.11878.
\newblock \doarXiv{2110.11878}

\bibitem[{{Soler} {et~al.}(2018){Soler}, {Bracco}, \& {Pon}}]{soler_2018}
{Soler}, J.~D., {Bracco}, A., \& {Pon}, A. 2018, \aap, 609, L3,
  \dodoi{10.1051/0004-6361/201732203}

\bibitem[{{Swiggum} {et~al.}(2022){Swiggum}, {Alves}, {D'Onghia}, {Benjamin},
  {Thulasidharan}, {Zucker}, {Poggio}, {Drimmel}, {Gallagher}, \&
  {Goodman}}]{swiggum_2022}
{Swiggum}, C., {Alves}, J., {D'Onghia}, E., {et~al.} 2022, arXiv e-prints,
  arXiv:2204.06003.
\newblock \doarXiv{2204.06003}

\bibitem[{{Taank} {et~al.}(2022){Taank}, {Marchal}, \& {Martin}}]{taank_2022}
{Taank}, M., {Marchal}, A., \& {Martin}, P. 2022, in preparation

\bibitem[{{Tetzlaff} {et~al.}(2013){Tetzlaff}, {Torres}, {Neuh{\"a}user}, \&
  {Hohle}}]{tetzlaff_2013}
{Tetzlaff}, N., {Torres}, G., {Neuh{\"a}user}, R., \& {Hohle}, M.~M. 2013,
  \mnras, 435, 879, \dodoi{10.1093/mnras/stt1358}

\bibitem[{{Tritsis} {et~al.}(2019){Tritsis}, {Federrath}, \&
  {Pavlidou}}]{tritsis_2019}
{Tritsis}, A., {Federrath}, C., \& {Pavlidou}, V. 2019, \apj, 873, 38,
  \dodoi{10.3847/1538-4357/ab037d}

\bibitem[{{van der Velden}(2020)}]{velden_2020}
{van der Velden}, E. 2020, The Journal of Open Source Software, 5, 2004,
  \dodoi{10.21105/joss.02004}

\bibitem[{{van der Walt} {et~al.}(2011){van der Walt}, {Colbert}, \&
  {Varoquaux}}]{van_der_walt_2011}
{van der Walt}, S., {Colbert}, S.~C., \& {Varoquaux}, G. 2011, CSE, 13, 22

\bibitem[{{van Woerden} {et~al.}(1999){van Woerden}, {Schwarz}, {Peletier},
  {Wakker}, \& {Kalberla}}]{van_Woerden_1999}
{van Woerden}, H., {Schwarz}, U.~J., {Peletier}, R.~F., {Wakker}, B.~P., \&
  {Kalberla}, P. M.~W. 1999, \nat, 400, 138, \dodoi{10.1038/22061}

\bibitem[{{Vergely} {et~al.}(2001){Vergely}, {Freire Ferrero}, {Siebert}, \&
  {Valette}}]{vergely_2001}
{Vergely}, J.~L., {Freire Ferrero}, R., {Siebert}, A., \& {Valette}, B. 2001,
  \aap, 366, 1016, \dodoi{10.1051/0004-6361:20010006}

\bibitem[{{Virtanen} {et~al.}(2020){Virtanen}, {Gommers}, {Oliphant},
  {Haberland}, {Reddy}, {Cournapeau}, {Burovski}, {Peterson}, {Weckesser},
  {Bright}, {van der Walt}, {Brett}, {Wilson}, {Jarrod Millman}, {Mayorov},
  {Nelson}, {Jones}, {Kern}, {Larson}, {Carey}, {Polat}, {Feng}, {Moore}, {Vand
  erPlas}, {Laxalde}, {Perktold}, {Cimrman}, {Henriksen}, {Quintero}, {Harris},
  {Archibald}, {Ribeiro}, {Pedregosa}, {van Mulbregt}, \&
  {Contributors}}]{SciPy-NMeth}
{Virtanen}, P., {Gommers}, R., {Oliphant}, T.~E., {et~al.} 2020, Nature
  Methods, 17, 261, \dodoi{https://doi.org/10.1038/s41592-019-0686-2}

\bibitem[{{Wakker} {et~al.}(1996){Wakker}, {Howk}, {Schwarz}, {van Woerden},
  {Beers}, {Wilhelm}, {Kalberla}, \& {Danly}}]{wakker_1996}
{Wakker}, B., {Howk}, C., {Schwarz}, U., {et~al.} 1996, \apj, 473, 834,
  \dodoi{10.1086/178196}

\bibitem[{{Wakker} {et~al.}(2003){Wakker}, {Savage}, {Sembach}, {Richter},
  {Meade}, {Jenkins}, {Shull}, {Ake}, {Blair}, {Dixon}, {Friedman}, {Green},
  {Green}, {Kruk}, {Moos}, {Murphy}, {Oegerle}, {Sahnow}, {Sonneborn},
  {Wilkinson}, \& {York}}]{wakker_2003}
{Wakker}, B.~P., {Savage}, B.~D., {Sembach}, K.~R., {et~al.} 2003, \apjs, 146,
  1, \dodoi{10.1086/346230}

\bibitem[{{Welsh}(1991)}]{welsh_1991}
{Welsh}, B.~Y. 1991, \apj, 373, 556, \dodoi{10.1086/170074}

\bibitem[{{Welsh} {et~al.}(1994){Welsh}, {Craig}, {Vedder}, \&
  {Vallerga}}]{welsh_1994}
{Welsh}, B.~Y., {Craig}, N., {Vedder}, P.~W., \& {Vallerga}, J.~V. 1994, \apj,
  437, 638, \dodoi{10.1086/175028}

\bibitem[{{Welsh} {et~al.}(2002){Welsh}, {Sallmen}, {Sfeir}, {Shelton}, \&
  {Lallement}}]{welsh_2002}
{Welsh}, B.~Y., {Sallmen}, S., {Sfeir}, D., {Shelton}, R.~L., \& {Lallement},
  R. 2002, \aap, 394, 691, \dodoi{10.1051/0004-6361:20021165}

\bibitem[{{Welsh} {et~al.}(1999){Welsh}, {Sfeir}, {Sirk}, \&
  {Lallement}}]{welsh_1999}
{Welsh}, B.~Y., {Sfeir}, D.~M., {Sirk}, M.~M., \& {Lallement}, R. 1999, \aap,
  352, 308

\bibitem[{Winkel {et~al.}(2016)Winkel, Kerp, Flöer, Kalberla, Ben~Bekhti,
  Keller, \& Lenz}]{winkel_effelsberg-bonn_2016}
Winkel, B., Kerp, J., Flöer, L., {et~al.} 2016, \aap, 585, A41,
  \dodoi{10.1051/0004-6361/201527007}

\bibitem[{{Wolfire} {et~al.}(1995){Wolfire}, {Hollenbach}, {McKee}, {Tielens},
  \& {Bakes}}]{wolfire_1995}
{Wolfire}, M.~G., {Hollenbach}, D., {McKee}, C.~F., {Tielens}, A.~G.~G.~M., \&
  {Bakes}, E.~L.~O. 1995, \apj, 443, 152, \dodoi{10.1086/175510}

\bibitem[{{Wolfire} {et~al.}(2003){Wolfire}, {McKee}, {Hollenbach}, \&
  {Tielens}}]{wolfire_2003}
{Wolfire}, M.~G., {McKee}, C.~F., {Hollenbach}, D., \& {Tielens}, A.~G.~G.~M.
  2003, \apj, 587, 278, \dodoi{10.1086/368016}

\bibitem[{{Zucker} {et~al.}(2019){Zucker}, {Speagle}, {Schlafly}, {Green},
  {Finkbeiner}, {Goodman}, \& {Alves}}]{zucker_2019}
{Zucker}, C., {Speagle}, J.~S., {Schlafly}, E.~F., {et~al.} 2019, \apj, 879,
  125, \dodoi{10.3847/1538-4357/ab2388}

\bibitem[{{Zucker} {et~al.}(2021){Zucker}, {Goodman}, {Alves}, {Bialy}, {Koch},
  {Speagle}, {Foley}, {Finkbeiner}, {Leike}, {En{\ss}lin}, {Peek}, \&
  {Edenhofer}}]{zucker_2021}
{Zucker}, C., {Goodman}, A., {Alves}, J., {et~al.} 2021, \apj, 919, 35,
  \dodoi{10.3847/1538-4357/ac1f96}

\bibitem[{{Zucker} {et~al.}(2022){Zucker}, {Goodman}, {Alves}, {Bialy},
  {Foley}, {Speagle}, {Gro{\ss}schedl}, {Finkbeiner}, {Burkert}, {Khimey}, \&
  {Swiggum}}]{zucker_2022}
{Zucker}, C., {Goodman}, A.~A., {Alves}, J., {et~al.} 2022, arXiv e-prints,
  arXiv:2201.05124.
\newblock \doarXiv{2201.05124}

\end{thebibliography}
\end{document}